\documentclass[preprintnumbers,aps,amsmath,amssymb,pra,twocolumn,showpacs,superscriptaddress]{revtex4-1}

\usepackage[dvipdfmx]{graphicx}
\usepackage{bm}
\usepackage{revsymb} 
\usepackage{amsthm} 
\usepackage{float}	
\usepackage{color}

\newcommand{\ham}{\mathcal{H}}

\newcommand{\ntg}{\notag \\}
\newcommand{\bra}[1]{\langle {#1} |}
\newcommand{\ket}[1]{| {#1} \rangle}
\newcommand{\Bra}[1]{\langle {#1} |}
\newcommand{\Ket}[1]{| {#1} \rangle}
\newcommand{\kb}[1]{\Ket{#1} \Bra{#1}}

\newcommand{\Eqref}[1]{Eq.~\eqref{#1}}
\newcommand{\Figref}[1]{FIG.~\ref{#1}}
\newcommand{\ctr}[1]{\cite{RefWorks:#1}}

\newcommand{\id}{\leavevmode\hbox{\small1\kern-3.3pt\normalsize1}}

\begin{document}

\title{Parallelizable adiabatic gate teleportation}

\author{Kosuke Nakago}
\affiliation{Department of Physics, Graduate School of Science, The University of Tokyo,\\ 7-3-1 Hongo, Bunkyo-ku, Tokyo, Japan 113-0033.}
\author{Michal Hajdu\v{s}ek}
\affiliation{Department of Physics, Graduate School of Science, The University of Tokyo,\\ 7-3-1 Hongo, Bunkyo-ku, Tokyo, Japan 113-0033.}
\affiliation{Singapore University of Technology and Design, 8 Somapah Road, Singapore 487372.}
\author{Shojun Nakayama}
\affiliation{Department of Physics, Graduate School of Science, The University of Tokyo,\\ 7-3-1 Hongo, Bunkyo-ku, Tokyo, Japan 113-0033.}
\affiliation{Institute for Interdisciplinary Information Sciences (IIIS), Tshighua University, FIT Building 1-208}
\author{Mio Murao}
\affiliation{Department of Physics, Graduate School of Science, The University of Tokyo,\\ 7-3-1 Hongo, Bunkyo-ku, Tokyo, Japan 113-0033.}
\affiliation{Institute for Nano Quantum Information Electronics, The University of Tokyo,\\ 4-6-1 Komaba, Meguro-ku, Tokyo, Japan 153-8505.}
\date{\today}

\begin{abstract}

To investigate how a temporally ordered gate sequence can be parallelized in adiabatic implementations of quantum computation, we modify adiabatic gate teleportation, a model of quantum computation proposed  by D. Bacon and S.~T. Flammia, Phys. Rev. Lett. {\bf 103}, 120504 (2009), to a form deterministically simulating parallelized gate teleportation, which is achievable only by postselection.    We introduce a {\it twisted Heisenberg-type interaction Hamiltonian}, a Heisenberg-type spin interaction where the coordinates of the second qubit are twisted according to a unitary gate.  We develop {\it parallelizable adiabatic gate teleportation} (PAGT) where a sequence of unitary gates is performed in a single step of the adiabatic process.  In PAGT, numeric calculations suggest the necessary time for the adiabatic evolution implementing a sequence of $L$ unitary gates increases at most as $O(L^5)$.  However, we show that it has the interesting property that it can map the temporal order of gates to the spatial order of interactions specified by the final Hamiltonian.  Using this property, we present a controlled-PAGT scheme to manipulate the order of gates by a control-qubit. In the controlled-PAGT scheme, two differently ordered sequential unitary gates $FG$ and $GF$ are coherently performed depending on the state of a control-qubit by simultaneously applying the twisted Heisenberg-type interaction Hamiltonians implementing unitary gates $F$ and $G$.   We investigate why the twisted Heisenberg-type interaction Hamiltonian allows PAGT.  We show that the twisted Heisenberg-type interaction Hamiltonian has an ability to perform a transposed unitary gate by just modifying the space ordering of the final Hamiltonian implementing a unitary gate in adiabatic gate teleportation. The dynamics generated by the time-reversed Hamiltonian represented by the transposed unitary gate enables deterministic simulation of a postselected event of parallelized gate teleportation in adiabatic implementation.

\end{abstract}
\pacs{03.67.Lx, 03.67.Ac, 75.10.Pq}
\maketitle

\section{Introduction}
The quantum circuit model is a standard model of quantum computation describing the relationship between input and output by a sequence of elementary gates. This model is widely used since it is universal, namely, any unitary operation can be represented by a sequence of the elementary gates, and quantum circuits have a good correspondence to logic circuits used in classical computation.  However, there is a restriction that elementary gates have to be performed in sequence determined by a partial order, without creating any loops in the circuit.  Since the partial order determines the causal structure of a gate sequence, only operations with definite temporal order can be performed, whereas such a restriction may not be necessary in quantum mechanics as pointed out in the work of Chiribella {\it et al.}~\cite{RefWorks:19}.

In \ctr{19}, an operation beyond temporally ordered quantum computation called quantum switch was investigated.  Quantum switch is a super-map which takes two different single-qubit unitary gates $F$ and $G$ as the input, and outputs a two-qubit controlled-unitary that coherently performs two differently ordered sequential unitary operations $FG$ and $GF$ depending on the state of a control-qubit.   It was proven that the quantum switch cannot be implemented within the quantum circuit model with a fixed temporal order, if each of $F$ and $G$ is allowed to be used only once. 

In this paper, we show that by implementing the quantum computation adiabatically, one can map the temporal order of gates to the spatial order of interactions, and that the temporal order can be manipulated by arranging the spatial order of the interactions Hamiltonian. To achieve this task, we modify adiabatic gate teleportation (AGT), which is a model of quantum computation proposed by Bacon and Flammia \cite{RefWorks:30}.  AGT is a hybrid of quantum circuit model and adiabatic quantum computation \cite{RefWorks:38}.   

AGT implements a unitary gate on an input state encoded in a degenerate ground state and transfers the state to an output state via a mediating qubit by adiabatically changing an initial interaction Hamiltonian encoding the unitary gate to a fixed final interaction Hamiltonian.  Since the input state can be chosen arbitrarily in AGT in contrast to standard adiabatic computation where input states are fixed, a sequence of unitary gates can be performed by sequentially combining each AGT process.  

Original gate teleportation \cite{GottesmanChuang} is a scheme which applies an arbitrary unitary gate to an unknown input state and transmits the resulting state to an output state using the quantum teleportation scheme \cite{Teleportation} with an entangled state encoding the unitary gate as a resource. A correction depending on the outcome of the Bell measurement on the input and a part of the resource state is required to deterministically perform gate teleportation. This correction depends on the unitary gate being teleported.  However if we consider the case of a particular Bell measurement outcome, the correction is not necessary.   In this case of a postselected event of the measurement outcome, we can select an input state after applying the unitary gate on an entangled state for preparing the resource state, so the temporal order may be interpreted to be distorted in postselected gate teleportation.   A certain type of quantum gates without definite temporal order (acausal gates) has been modeled based on postselected quantum teleportation \cite{P-CTC:BS, RefWorks:75}.  These gates are referred to as closed timelike curves via quantum postselection (p-CTCs, or also known as BSS-CTCs) and their properties have been recently intensively investigated \cite{RefWorks:31, RefWorks:33, RefWorks:34,RefWorks:35,RefWorks:37}.  

In spite of its name, AGT is not a scheme based on quantum teleportation that requires entanglement assisted local operations and classical communications (LOCC).  AGT {\it deterministically} implements a map that is achieved by postselected gate teleportation of an arbitrary unitary gate by using adiabatic dynamics of a multi-qubit ground state under a time-dependent interaction Hamiltonian. This association of deterministically ``simulating'' a postselection event of a measurement using adiabatic dynamics is useful for analyzing the role of temporal order and parallelizability in quantum computation. 

A sequence of unitary gates can be {\it probabilistically} implemented in a {\it parallel} manner by combining multiple gate teleportations. Each unitary gate can be simultaneously performed on different resource states and all the measurements can be simultaneously performed in this case.  However, the original AGT scheme proposed by \cite{RefWorks:30} cannot simulate this parallelizable property of postselected gate teleportation.  A sequence of unitary gates has to be implemented by {\it sequentially} applying the AGT scheme in time for each gate following the temporal order of the gate sequence.  

In this paper, we modify the Hamiltonian of the original AGT scheme to a Heisenberg-type spin interaction in order to simulate parallelized gate teleportation.  With this modification, we introduce a {\it twisted Heisenberg-type interaction Hamiltonian} as a resource to implement a unitary gate. Using the twisted Heisenberg-type interaction Hamiltonian, we develop parallelizable adiabatic gate teleportation (PAGT), a scheme to perform a sequence of unitary gates in one adiabatic step from the initial Hamiltonian to the final Hamiltonian.  In PAGT, only the temporal order of gate operations is mapped to spatial order of interactions in the final Hamiltonian, whereas information of which gates to be applied is encoded in twisted angles of the initial twisted Heisenberg-type interaction Hamiltonian. 

All twisted Heisenberg-type interaction Hamiltonians are simultaneously applied, though the rate of the adiabatic evolution must take into account the decreasing gap of the total Hamiltonian. PAGT does not contribute to speed-up performing a sequence of gates by parallelizing them. However it enables the control of temporal order of gate operations in the time domain. Using this property, we present the controlled-PAGT scheme that performs controlled-unitary operations implemented by the quantum switch. 

We also investigate why the twisted Heisenberg-type interaction Hamiltonian allows PAGT.  We show that the twisted Heisenberg-type interaction Hamiltonian also allows simulation of a transposed unitary gate by using the similar resource implementing the original non-transposed unitary gate but the spatial order of the interaction is changed. We show that this is not possible in general in the AGT scheme using a non-Heisenberg type interaction Hamiltonian.  We analyze this ability to perform a transposed unitary gate and its relationship to implementability of dynamics generated by the time-reversed Hamiltonian.

This paper is organized as follows. We present preliminaries to review the adiabatic theorem and adiabatic gate teleportation in Section \ref{adiabatic theorem} and Section \ref{subsec:agt}, respectively. In Section \ref{sec:gateham}, we introduce the twisted Heisenberg-type interaction Hamiltonian and analyze its properties. In Section \ref{PAGT} and Section \ref{CPAGT}, we present our main results, the PAGT scheme and the controlled-PAGT scheme, respectively. We compare implementability of dynamics generated by the time-reversed Hamiltonian in AGT and PAGT, and analyze parallelizability in PAGT in Section \ref{TRHdynamics}.   We summarize our results in Section \ref{conclusion}.

\section{Preliminaries}
\label{preliminaries}

\subsection{Adiabatic theorem}
\label{adiabatic theorem}
Adiabatic theorem is a well-known concept concerning quantum dynamics and was first proposed in 1928 by Born and Fock~\cite{originalAGT,Kato}. 
In \cite{originalAGT} (originally written in German), the theorem was stated as ``A physical system remains in its instantaneous eigenstate under the time-evolution given by a time-dependent Hamiltonian if there is a gap between the eigenvalue and the rest of Hamiltonian's eigenvalues and the Hamiltonian changes slowly enough (adiabatic evolution)".  While its proof was first provided under the so-called assumption of the {\it adiabatic approximation}, Gell-Mann and Low~\cite{GL} have constructed an adiabatic theorem for infinitely slow adiabatic evolution, namely, the limit $T\rightarrow \infty$ where $T$ is the total evolution time of the time-dependent Hamiltonian dynamics and they have proved the theorem without the adiabatic approximation assumption.

Adiabatic quantum computation (AQC)~\ctr{38} is a model of quantum computation, which is originally proposed for exploiting the adiabatic theorem to solve the Boolean Satisfiability Problem (SAT).   To employ adiabatic evolutions for solving computational problems, a new type of adiabatic theorem is required since the formulation based on the theorem that holds only for infinite runtime $T$ is not useful for analyzing the time required for computation.  Thus adiabatic theorems \cite{JRS,Wiebe,Lidar} referring to the trade-off between the runtime $T$ and the error $\epsilon$ from the ideal result have been proposed.

Let us consider a time-dependent Hamiltonian $H(\tau)$, where $\tau=t/T$ is the rescaled time, $\tau\in[0,1]$. We denote the instantaneous eigenstate and the corresponding eigenenergy of $H(\tau)$ by $\ket{E_n(\tau)}$ and $E_n(\tau)$, respectively, where $n=0$ represents the ground state and $n=1,2,\cdots$ represents the $n$-th excited state. In adiabatic computation, it is usual to consider an interpolation between two time-independent Hamiltonians, an initial Hamiltonian $H_{\text{ini}}$ and a final one $H_{\text{fin}}$.  In this work, we consider a particular interpolation given by $H(\tau)=(1-s(\tau))H_{\text{ini}}+s(\tau)H_{\text{fin}}$, where $s(0)=0$ and $s(1)=1$. Lidar {\it et al.}~\cite{Lidar} have shown that the following conditions need to be fulfilled in order for the adiabatic theorem to hold.

\textbf{Assumption 1.} $H(\tau)$ is a one-parameter bounded Hamiltonian of an $n$-partite system defined on a tensor product space $\mathcal{H}^{\otimes n}$. A nonnegative real parameter $\gamma$ is the distance to a pole or a branch point of $H(\tau)$ that is closest to the real $\tau$-axis of an interval $[0,1]$ in the complex $\tau$-plane.

\textbf{Assumption 2.} For any number less than a fixed number $N$, it is possible to set the number of derivatives of $H(\tau)$ with respect to $\tau$ to zero at the initial and final times.

\textbf{Assumption 3.} The final state $\ket{E_0 (1)}$ is non-degenerate.

Before stating the adiabatic theorem of \cite{Lidar}, we introduce some notations.  The vector norm is defined in the usual way as $\|v\|=\sqrt{\bra{v}v\rangle}$.  For an operator $A$, its operator norm is defined as 
\begin{equation*}
	\| A \| := \max_{\ket{v}}|\bra{v} A \ket{v}|, 
\end{equation*}
where  $\ket{v}$ is a normalized vector.   Quantity indicating the largest ``curvature'' of the time-dependent Hamiltonian for  $0\leq \tau \leq 1$ is defined by
\begin{equation*}
	\xi : = \sup_{0\leq \tau \leq 1} \| \partial_{\tau} H(\tau)\|.
\end{equation*}
The minimum gap between the ground state(s) and the first excited state(s) over $0\leq \tau \leq 1$ is denoted as
\begin{equation*}
\mathcal{G} = \min_{0\leq \tau\leq 1}\left(E_1(\tau) - E_0(\tau)\right).
\end{equation*}

In adiabatic computation, the system is usually initialized in the ground state of the initial Hamiltonian $H_{\text{ini}}$. One of the main quantities of interest is the distance $\delta$ between the state of the system $\ket{\psi(\tau)}$ and the ground state $\ket{E_0(\tau)}$ at the end of the evolution. This distance can be computed using the vector norm as $\delta=\|\ket{\psi(1)} - e^{i\chi}\ket{E_0(1)}\|$, where $\chi\in\mathbb{R}$ is the dynamical phase. In particular, the trade-off relation between the total evolution time $T$ and the adiabatic approximation error $\delta$ is of crucial importance. Lidar {\it et al.}~have shown in \cite{Lidar} that under some circumstances the adiabatic approximation can be made exponentially accurate in $N$, as stated in the following theorem (Theorem 1 in \cite{Lidar}).

{\bf Theorem.} {\it If assumptions 1-3 hold and the first $N+1$ derivatives of the Hamiltonian $H(\tau)$ all vanish at $\tau=0$ and $\tau=1$}, then the runtime $T$ that scales as
\begin{equation}\label{totaltime}
	T = \frac{q}{\gamma}N\frac{\xi^2}{\mathcal{G}^3},
\end{equation}
{\it where the ``time dilation" $q>1$ is a free parameter, gives an adiabatic approximation error satisfying}
\begin{equation*}
 	\delta \leq (N + 1)^{\gamma + 1} q^{-N}.
\end{equation*}

This theorem gives the condition for the runtime $T$ of an adiabatic evolution so that the error is exponentially small in terms of the number of vanishing derivatives.  The following corollary will be useful in our paper (Corollary 2 in \cite{Lidar}).

{\bf Corollary.} {\it Under the assumptions of the Theorem above, the following scaling of runtime $T$,
\begin{equation}\label{totaltime2}
	T = \epsilon^{-1/N}\frac{N(N+1)^{\frac{\gamma+1}{N}}}{\gamma}\frac{\xi^2}{\mathcal{G}^3},
\end{equation}
{\it gives an adiabatic error satisfying}
\begin{equation*}
	\delta \leq \epsilon.
\end{equation*}
This corollary demonstrates that the runtime $T$ is insensitive to the adiabatic error, as it depends only on its $N$-th root.}

\subsection{Adiabatic gate teleportation} \label{subsec:agt}
In this subsection, we review adiabatic gate teleportation (AGT) proposed in \cite{RefWorks:30}. We consider systems consisting of only qubits in this paper, but one can generalize the scheme to general qudit (quantum $d$-level) systems. 
We use $X,Y$ and $Z$ to represent the Pauli operators, generators of SU(2), and use $I$ to represent an identity operator on a qubit system represented by a Hilbert space denoted by $\mathcal{H}=\mathbb{C}^2$.

We use subscript $i$ to denote the local Hilbert space on which a state is defined or on which an operator acts nontrivially.  We also drop the tensor product symbol. For example, $\ket{\phi}_1 U_3 \ket{\Phi^+}_{23} \in \ham_1 \otimes \ham_2 \otimes \ham_3$ is a 3-qubit state where $U_3 \ket{\Phi^+}_{23} $ is a simplified representation of $(I_2 \otimes U_3) \ket{\Phi^+}_{23}$.   If we want to represent an operator without specifying the Hilbert space on which the operator acts, we omit the subscript and just denote the operator as $A$.

Consider a three-qubit system $\ham_1 \otimes \ham_2 \otimes \ham_3$. AGT aims to obtain the output state $U\ket{\phi}_3 \in \ham_3$ where $U$ denotes a single-qubit unitary gate for an arbitrary input state $\ket{\phi}_1 \in \ham_1$ by adiabatically evolving the system under a time-dependent Hamiltonian.  The system $\ham_2$ acts as a mediator for transferring the state. 

We introduce a two-qubit {\it twisted interaction Hamiltonian} ${{\tilde H}_{ij}}^U$ on $\ham_i \otimes \ham_j$ defined as
\begin{equation}
	{\tilde H}_{ij}^U := -\omega  U_j ( X_i X_j  + Z_i Z_j ) U_j^\dagger. \label{AGTinteration}
\end{equation}
In case $U_j=I_j$, we denote ${\tilde H}_{ij}^U$ simply as ${\tilde H}_{ij}$.  
We consider an initial Hamiltonian ${\tilde H}_{\rm ini}^U$ on $\ham_1 \otimes \ham_2 \otimes \ham_3$ given by
\begin{equation*}
	{\tilde H}_{{\rm ini}}^U= {\tilde H}_{23}^U,
\end{equation*}
and a final Hamiltonian ${\tilde H}_{\rm fin}$ on $\ham_1 \otimes \ham_2 \otimes \ham_3$ given by 
\begin{equation*}
	{\tilde H}_{\rm fin} ={\tilde H}_{12},
\end{equation*}
where $\omega >0$ for both Hamiltonians.

We prepare an initial state in $\ham_1 \otimes \ham_2 \otimes \ham_3$ as $\ket{\phi}_1 U_3 \ket{\Phi^+}_{23}$, 
where $\ket{\phi}$ is an arbitrary input state and $\ket{\Phi^+}=(\ket{00}+\ket{11} )/ \sqrt{2}$ is a two-qubit maximally entangled state. 
It is a ground state of the initial Hamiltonian ${\tilde H}_{{\rm ini}}^U$ on $\ham_1 \otimes \ham_2 \otimes \ham_3$ because the following holds true,  
\begin{eqnarray*}
	(X_2 U_3 X_3 U_3^\dagger) U_3 \ket{\Phi^+}_{23} &=& U_3 \ket{\Phi^+}_{23}, \\
	(Z_2 U_3 Z_3 U_3^\dagger) U_3 \ket{\Phi^+}_{23} &=& U_3 \ket{\Phi^+}_{23}.
\end{eqnarray*}
The initial Hamiltonian acts trivially on $\ham_1$ which means the ground state is 2-fold degenerate.  We can choose an arbitrary input state $\ket{\phi}_1 \in \ham_1$, but the state of the second and third qubits is fixed. Similarly, the ground state of the final Hamiltonian ${\tilde H}_{{\rm fin}}$ is given by $\ket{\Phi^+}_{12} \ket{\phi^\prime}_3$ where $ \ket{\phi^\prime}_3$ is an arbitrary state in $\mathcal{H}_3$.

The total Hamiltonian on $\ham_1 \otimes \ham_2 \otimes \ham_3$ is described by 
\begin{equation}
	H_{\rm AGT}^U (\tau) = \left(1-s(\tau)\right){\tilde H}_{\rm ini}^U + s(\tau) {\tilde H}_{\rm fin}, \label{ATham}
\end{equation}
where the interpolation function satisfies $s(0)=0$ and $s(1)=1$.  To show that adiabatic evolution by $H_{\rm AGT}^U (\tau)$ transforms an initial state  $\ket{\phi}_1 U_3\ket{\Phi^+}_{23}$ to a final state $\ket{\Phi^+}_{12} U_3 \ket{\phi}_3$ for arbitrary $\ket{\phi}_1$, we define logical operators $L_X^{U_3}:=X_1 X_2 U_3 X_3  U_3^\dagger$ and  $L_Z^{U_3}:= Z_1 Z_3 U_3 Z_3  U_3^\dagger$, both on $\ham_1 \otimes \ham_2 \otimes \ham_3$.  These logical operators encode information of an arbitrary input state $\ket{\phi}_1$, and if the logical space spanned by the logical operators is preserved during adiabatic evolution, then we can show $ \ket{\phi^\prime}_3 = U_3 \ket{\phi}$.

First we consider the case when $U=I$ which corresponds to {\it adiabatic teleportation}.  The total Hamiltonian $H_{\rm AT} (\tau):= H_{\rm AGT}^I (\tau)$ has two degenerate ground states for all $\tau \in [0,1]$ and no level-crossing between the degenerate ground states and the two degenerate excited states. The gap between the ground states and the excited states reaches its minimum at $s(\tau)=1/2$.  In this case, the logical operators are given by $L_X:=X_1 X_2 X_3 $ and  $L_Z:= Z_1 Z_3 Z_3$ and they commute with the total Hamiltonian $H_{\rm AT} (\tau)$,
\begin{equation}
	[L_X, H_{\rm AT} (\tau)]= [L_Z, H_{\rm AT} (\tau)]=0, \label{atcommutation}
\end{equation}
for all $\tau \in [0,1]$.  Faithful transformation of an input state $\ket{\phi}$ from $\mathcal{H}_1$  to $\mathcal{H}_3$ is guaranteed by the adiabatic theorem since the adiabatic evolution proceeds inside the logical subspace.   Thus by adiabatically dragging the initial Hamiltonian to the final one, the initial state $\ket{\phi}_1 \ket{\Phi^+}_{23}$ is transformed to $\ket{\Phi^+}_{12} \ket{\phi}_3$, since
\begin{align}
 \bra{\phi}X\ket{\phi} &= \bra{\phi}\bra{\Phi^+} L_X \ket{\phi}\ket{\Phi^+} \nonumber \\
&= \bra{\Phi^+}\bra{\phi'} L_X \ket{\Phi^+}\ket{\phi'} = \bra{\phi'}X\ket{\phi'}, \\
\bra{\phi}Z\ket{\phi} &= \bra{\phi'}Z\ket{\phi'}. 
\end{align}

It should be noted that the adiabatic theorem of \cite{Lidar} applies to non-degenerate ground states and the Hamiltonians in the AGT scheme and throughout this paper have a doubly degenerate ground state. We know that the individual logical subspaces are invariant under the adiabatic evolution due to the commutation relation Eq.~(\ref{atcommutation}). Also, we are interested in the evolution of an encoded qubit given by the intersection of these two subspaces which is a one-dimensional subspace. Therefore the ground state is effectively non-degenerate and theorem of \cite{Lidar} applies.

For $U \neq I$, the total Hamiltonian $H_{AGT}^U(\tau)$ is related to the previous case by
\begin{equation*}
	H_{\rm AGT}^U (\tau) = U_3 H_{\rm AT} (\tau) U_3^\dagger, \label{AGTham}
\end{equation*}
since $U_3 {\tilde H}_{\rm fin} U_3^\dagger = {\tilde H}_{\rm fin}$.  As this is just a unitary conjugation, the spectrum of $H_{\rm AGT}^U (\tau)$ remains unchanged.  By defining unitary conjugated logical operators as $L_X^{U_3}:=X_1 X_2 (U_3 X_3  U_3^\dagger)$ and  $L_Z^{U_3}:= Z_1 Z_3 (U_3 X_3  U_3^\dagger)$, both on $\ham_1 \otimes \ham_2 \otimes \ham_3$, the commutation relations satisfy
\begin{equation}
	[L_X^{U_3}, H_{\rm AGT}^U (\tau)]= [L_Z^{U_3}, H_{\rm AGT}^U (\tau)]=0,  
\end{equation}
for all $\tau \in [0,1]$.  This shows that the initial state $\ket{\phi}$ on the first qubit is deterministically teleported to third qubit and the unitary gate $U$ is applied to it during the adiabatic evolution.

\begin{figure}[htb]
	\begin{center}
	\includegraphics[clip, width=0.7\columnwidth]{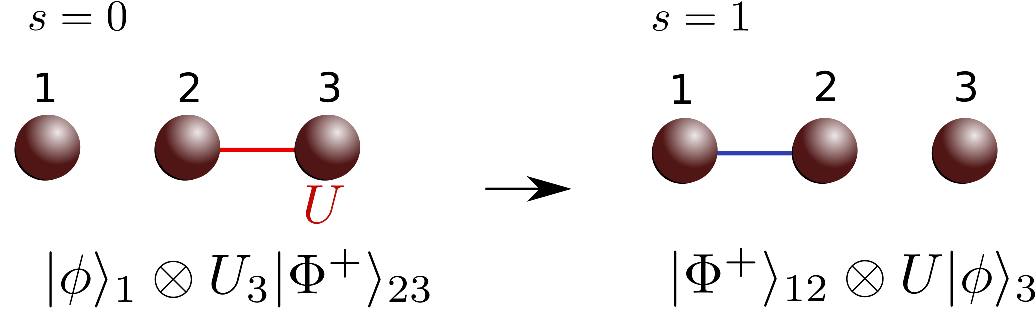}
	\end{center}
 	\caption{(Color online) A schematic how AGT works. The picture on the left represents the initial state of the adiabatic evolution. The red line is the interaction between the second and third qubits due to the initial Hamiltonian. The picture on the right is the output state at the end of the evolution. The blue line represents the interaction between the first and the second qubits.}\label{AGTcircuit}
\end{figure}

\subsection{Quantum switch} \label{sec:qs}

Under the definition of a quantum circuit, only the operations with definite temporal order can be performed, whereas such a restriction may not be necessary in quantum mechanics as pointed out in \cite{RefWorks:19}.  As an example of acausal operations, we give a brief review of the {\it quantum switch} introduced in \cite{RefWorks:19}. 

Quantum switch is a mechanism that controls the order of gates depending on the state of the control-qubit.  We consider a two-qubit system consisting of a control-qubit and a target-qubit, described by $\ham_C \otimes \ham_T$. We also consider two single-qubit unitary gates on $\ham_T$, $F$ and $G$. 
Quantum switch is a supermap  (a map taking a map as an input and another map as an output) that takes $F$ and $G$ as inputs and returns a unitary operation given by
\begin{equation*}
	U_{\rm QS}^{F,G}: =\ket{0}\bra{0}_C \otimes (GF)_T + \ket{1}\bra{1}_C \otimes (FG)_T,
\end{equation*}
 namely, the supermap of quantum switch $QS(F,G)$ is defined by
\begin{equation*}
	QS(F,G) := U_{\rm QS}^{F,G}.
\end{equation*}
Note that the definitions of a unitary operation $U_{\rm QS}^{F,G}$ and the supermap $QS(F,G)$ represents different levels of operations, one is a unitary map and the other is a supermap.    Applying $U_{\rm QS}^{F,G}$ on a state given by $\ket{0}_C \ket{\phi_0}_T + \ket{1}_C \ket{\phi_1}_T \in \mathcal{H}_C \otimes \mathcal{H}_T$ where $\ket{\phi_{0}}, \ket{\phi_{1}}$ are arbitrary states satisfying $\langle \phi_0 \ket{\phi_0}+\langle \phi_1 \ket{\phi_1}=1$, we have 
\begin{align}
	U_{\rm QS}^{F,G}&\left( \ket{0}_C \ket{\phi_0}_T + \ket{1}_C \ket{\phi_1}_T \right) \nonumber\\
	&=\ket{0}_C GF\ket{\phi_0}_T + \ket{1}_C  FG\ket{\phi_1}_T. 
\label{QS}
\end{align}
When the control-qubit state is in $\ket{0}$, $F$ acts first on the target-qubit followed by $G$. If the control-qubit state in $\ket{1}$, the order of operations changes so that $G$ acts first followed by $F$.  Note that this quantum control preserves coherence between $\ket{0}_C$ and $\ket{1}_C$.

\begin{figure}[h]
	\begin{center}
   		\includegraphics[clip, width=0.7\columnwidth]{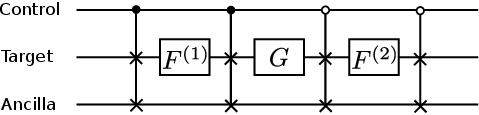}
  	\end{center}
  	\caption{ A quantum circuit implementing the quantum switch operation using multiple calls of the same oracle $F$ denoted by $F^{(1)}$ and $F^{(2)}$. 
The SWAP operation conditioned on the control-qubit state represented by a black circle is the controlled-SWAP (CSWAP) gate, it swaps the states of the target and ancilla qubits when the control-qubit is in $\ket{1}$. The SWAP operation conditioned on the control-qubit state represented by a white circle swaps the states of the target and ancilla when the control-qubit is in $\ket{0}$.}
  	\label{fig:QSbyQC}
\end{figure}

In the original paper \cite{RefWorks:19}, $F$ and $G$ are referred to as oracles. Similarly to the no-cloning theorem of unknown quantum states, the no-cloning theorem of oracles exists \ctr{69}.  The no-cloning theorem of oracles asserts that we cannot perform exactly the same unitary operation implemented by an oracle twice if we are only allowed to use the oracle once. We consider $F$ and $G$ as oracles where the identity of the unitary gates are unknown for now since we want to universally construct a supermap representing quantum switch independent of the identity of the gates $F$ and $G$.  It was shown in \ctr{19} that it is impossible to implement quantum switch by using only one call of each oracle $F, G$ in the quantum circuit model. 
The output state given by \Eqref{QS} can be produced, but multiple calls of at least one of the oracles are required (see \Figref{fig:QSbyQC}).  In the quantum circuit of \Figref{fig:QSbyQC}, oracles $F^{(1)}$ (the first call of oracle $F$), $G$, $F^{(2)}$ (the second call of oracle $F$) have to be arranged in a definite temporal ordering.

Even though the quantum switch cannot be implemented in the quantum circuit model, the assumption of {\it no definite causal structure} itself is not prohibited by the axioms of quantum mechanics. It was also suggested that if one can somehow implement {\it superposition of a quantum wire} in the quantum circuit, which is a wire controlled on the state of the control-qubit in the quantum circuit, we can implement quantum switch. 

Several studies have focused on the computational power of the quantum switch \cite{RefWorks:23, RefWorks:22, Araujo:2014_PRL,Procopio:2014, RefWorks:123}.
The definition of the quantum switch can be generalized so that $F$ and $G$ are quantum channels, i.e. completely positive trace-preserving (CPTP) maps. 
It has been proven that a quantum circuit augmented with the generalized quantum switch can achieve a channel discrimination task which is impossible in the standard quantum circuit model \cite{RefWorks:23}. The resources required for the order controlled operation implemented by the quantum switch and the standard quantum circuit are compared in \cite{RefWorks:22}.  Computational advantage of a quantum circuit augmented with the quantum switch has been proposed in \cite{Araujo:2014_PRL} and experimentally demonstrated in \cite{Procopio:2014}. The notion of quantum switch is generalized to a supermap in order to implement a $n$-SWITCH gate, which is a mechanism performing coherently superposed $n!$ differently ordered operations depending on the state of the control system in \cite{RefWorks:123}.

\section{Twisted Heisenberg-type interaction Hamiltonian} 
\label{sec:gateham}
\subsection{Definition}

To implement a sequence of unitary gates in AGT, a concatenation of adiabatic evolutions is necessary.  This concatenation introduces definite temporal ordering of the applied unitary gates. To remove the temporal ordering of a sequence of unitary gates and to simulate parallelized gate teleportation in an adiabatic manner, we modify the twisted interaction Hamiltonian used in AGT.

First, we present a modified Hamiltonian to implement a single unitary gate $U$ using the adiabatic evolution. This new twisted interaction Hamiltonian $H_{ij}^U$ on $\ham_i \otimes \ham_j$ depending on $U$ is defined as
\begin{equation*}
	{H_{ij}^U} : =-\omega  U_j ( X_i  X_j - Y_i Y_j + Z_iZ_j) U_j^\dagger.
\end{equation*}
Note that the operation denoted by $U$ acts on $\mathcal{H}_j$, the second qubit of $\mathcal{H}_i \otimes \mathcal{H}_j$ in our notation. Similarly to the case of AGT, we denote the case of $U=I$ by ${H_{ij}}$. The new twisted interaction Hamiltonian ${H_{ij}^U}$ contains an extra two-body interaction term $-Y_i Y_j$ compared to the twisted interaction Hamiltonian of the AGT scheme ${\tilde H}_{ij}^U$ given by Eq.~(\ref{AGTinteration}).   ${H_{ij}^U}$ represents a Heisenberg-type interaction where the $Z$-axis of system $\ham_j$ is twisted by a unitary gate $U_j$ and we refer to this type of Hamiltonian as {\it twisted Heisenberg-type interaction Hamiltonian}.  We note that the minus sign of the $-Y_i Y_j$ term is an important characteristic of this interaction, and is useful for simulating parallelized gate teleportation.  

The corresponding total Hamiltonian $H_{{\rm PAGT}}^U(\tau)$ on $\ham_1 \otimes \ham_2 \otimes \ham_3$ whose adiabatic evolution implements $U$ is defined as
\begin{equation}
	H_{{\rm PAGT}}^U(\tau)= (1-s(\tau)) H_{\rm ini}^U + s(\tau) H_{\rm fin}, \label{AGThamtotal} 
\end{equation}
where $H_{\rm ini}^U $ on $\ham_1 \otimes \ham_2 \otimes \ham_3$ is given by $H_{\rm ini}^U = H_{23}^U$, and $H_{\rm fin}$ on $\ham_1 \otimes \ham_2 \otimes \ham_3$ is given by $H_{\rm fin} = H_{12}$. We refer to the scheme of adiabatically implementing $U$ using $H_{{\rm PAGT}}^U(\tau)$ similarly to AGT as {\it parallelizable adiabatic gate teleportation} (PAGT).   This modification of the twisted interaction Hamiltonian in PAGT does not change the ground state of the initial and final Hamiltonians from the ones of AGT, while it changes the energy gap between the degenerate ground states and the degenerate first excited states. 

For $U=I$, the new total Hamiltonian $H_{{\rm PAGT}}^I (\tau)$ achieves adiabatic teleportation similarly to the original total Hamiltonian $H_{\rm AT} (\tau)=H_{\rm AGT}^I (\tau)$ which we will show in the next section.  Note that adding the $YY$ term does not affect the validity of adiabatic teleportation as already pointed out in the original paper of AGT \cite{RefWorks:30}. 

The twisted Heisenberg-type interaction Hamiltonian $H_{ij}^U$ can be represented in terms of the projector on the maximally entangled states $\ket{\Phi^+} \bra{\Phi^+}$ as
\begin{equation}
	H_{ij}^U = -\omega U_j ( 4 \ket{\Phi^+}\bra{\Phi^+}_{ij} - I_{ij} ) U^\dagger_j,   \label{gateham2}
\end{equation}
where $I_{ij}$ denotes an identity operator on $\ham_i \otimes \ham_j$.    
The ground state of $H_{ij}^U$ is given by $U_j \ket{\Phi^+}_{ij}$.  

Although we consider only qubit systems in this paper, the definition of twisted Hamiltonian can be generalized to qudit systems by 
\begin{equation}
	H_{ij}^U:= -\omega U_j ( d^2 \ket{\Phi_d^+}\bra{\Phi_d^+}_{ij} - I_i  I_j ) U^\dagger_j , \label{gengateham}
\end{equation}
where $\ket{\Phi_d ^+}:= 1/\sqrt{d} \sum_{i=0}^{d-1} \ket{ii}$ is the maximally entangled state on a $d$-dimensional system. 
The twisted Heisenberg-type interaction Hamiltonian has the following properties,
\begin{eqnarray}
H_{ij}^U &=&  U_j H_{ij} U^\dagger_j, \label{twist1}\\
H_{ij}^{U^T} &=& -\omega ( d^2 U^T_j\ket{\Phi_d^+}\bra{\Phi_d^+}_{ij} U_j^* - I_i \ I_j ) \nonumber\\
&=& -\omega  (d^2 U_i\ket{\Phi_d^+}\bra{\Phi_d^+}_{ij} U^\dagger_i - I_i  I_j ) \nonumber\\
&=&  U_i H_{ij}  U^{ \dagger}_i = H_{ji}^{U} , \label{twist2}
\end{eqnarray}
where we have used the property of the maximally entangled states 
\begin{equation*}
	U_j\ket{\Phi_d^+}_{ij}=U^T_i\ket{\Phi_d^+}_{ij}.
\end{equation*}
Similarly, $U_i^*  U_j\ket{\Phi_d^+}_{ij}=\ket{\Phi_d^+}_{ij}$ holds, therefore
\begin{equation}
	( U^*_i  U_j ) H_{ij} ( U^{*}_i  U_j )^\dagger =  H_{ij}. \label{twist3}
\end{equation}
These properties are extensively used in the rest of the paper.

Following assumptions need to be satisfied in order to implement $U$ using $H_{ij}^{U}$ in PAGT,
\begin{enumerate}
	\item The twisted Heisenberg-type interaction Hamiltonian $H_{ij}^{U}$ is implementable;
	\item The ground state of the twisted Heisenberg-type interaction Hamiltonian $U_3 \ket{\Phi^+}_{23}$ can be prepared in $\mathcal{H}_2 \otimes  \mathcal{H}_3$;
	\item Switching off the initial Hamiltonian and switching on the final Hamiltonian can be gradually done at the same time.
\end{enumerate}
We assume that these conditions are satisfied, and we consider the ideal case where no noise from the outside environment acts on the system in the rest of paper.

\subsection{Implementable unitary gates using a twisted Heisenberg-type interaction Hamiltonian}\label{power}

We show that  $U^{T}, U^*$ and $U^\dagger$ are implementable in addition to $U$ using the same twisted Heisenberg-type interaction Hamiltonian $H^U$ in PAGT by changing the spatial order of interactions and/or exchanging the initial and final Hamiltonians of the total Hamiltonian for adiabatic evolution.  In gate teleportation, $U^{T}, U^*$ and $U^\dagger$ are also implementable by using the resource state encoding $U$, but only probabilistically.  The PAGT scheme using the Heisenberg-type interaction deterministically replicates this aspect of gate teleportation, while the AGT scheme cannot implement $U^{T}$ in addition to $U$ in general as we will show in Section \ref{TRHdynamics}.

Using Eq.~(\ref{twist2}), a twisted Heisenberg-type interaction Hamiltonian $H_{ij}^U$ that implements $U$ can be transformed to a twisted Heisenberg-type interaction Hamiltonian implementing $U^T$ by just changing the spatial order of interaction $(i,j) \rightarrow (j,i)$ of qubits $i$ and $j$.  Thus, the initial Hamiltonian for implementing  $U^T$ can be written as $H_{\rm ini}^{U^T} = H_{32}^U$. By choosing an initial state for the adiabatic evolution as $\ket{\phi}_1 U_3^T \ket{\Phi^+}_{23} = \ket{\phi}_1 U_2 \ket{\Phi^+}_{23}$ and applying the total Hamiltonian given by
\begin{eqnarray*}
	H^{U^T}_{\rm PAGT} (\tau)&=&(1-s(\tau)) H_{\rm ini}^{U^T} + s(\tau) H_{\rm fin} \nonumber \\
	&=& (1-s(\tau)) H_{32}^U + s(\tau) H_{12},
\end{eqnarray*}
the final state of adiabatic evolution is given by $U^T \ket{\phi}_3$.  

\begin{figure}[hbtp]
        \begin{center}
            \includegraphics[clip, width=0.7\columnwidth]{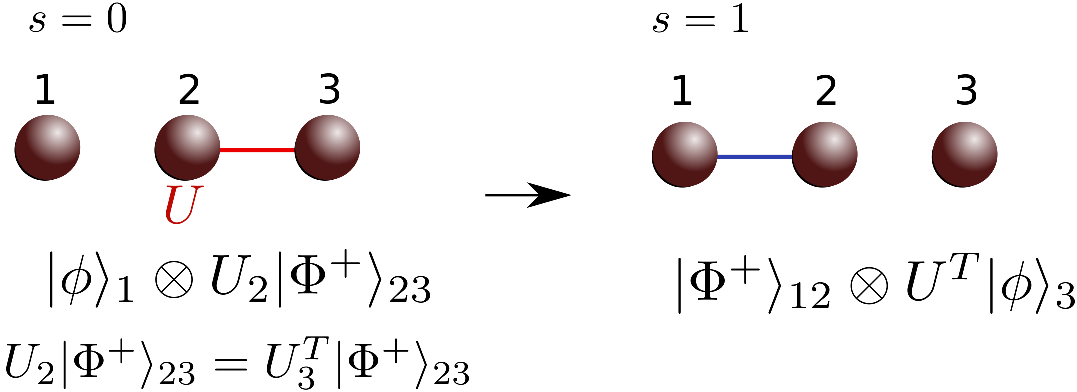}
        \end{center}
           \caption{Implementation of transposed unitary gate $U^T$.  The total Hamiltonian is given by $H^{U^T}_{\rm PAGT} (\tau)=(1-s(\tau)) H_{\rm ini}^{U^T} + s(\tau) H_{\rm fin} = (1-s(\tau)) H_{32}^U + s(\tau) H_{12}$.} 
	        \label{AGTinitial}
\end{figure}

To implement $U^*$,  we employ the relationship given by Eq.~(\ref{twist3}). We modify the initial and final Hamiltonian of the total Hamiltonian for adiabatic evolution as
\begin{equation}
	H^{U^*}_{\rm PAGT'} (\tau)=(1-s(\tau)) H_{\rm ini'} + s(\tau) H_{\rm fin'}^{U^*}, \label{AGTconj} 
\end{equation}
where $H_{\rm ini'}=H_{23}$ and $H_{\rm fin'}^{U^*} = H_{12}^{U}$.  Note that the $U$-dependence of the twisted Heisenberg-type interaction  Hamiltonian $H_{ij}^U$ appears in the final Hamiltonian in contrast to the case of the total Hamiltonian denoted by $H^{U}_{\rm PAGT} (\tau)$. The subscript of the total Hamiltonian ``${\rm PAGT'}$'' indicates this property (encoding information about $U$ in the final Hamiltonian).  Since $H^{U^*}_{\rm PAGT'} (\tau) = ( U_2  U_3^*) H^{I}_{\rm PAGT} (\tau)  (U_2  U_3^* )^\dagger$ holds due to Eq.~(\ref{twist3}), by choosing the initial state to be $\ket{\phi}_1 ( U_2 \otimes U_3^*) \ket{\Phi^+}_{23} = \ket{\phi}_1 \ket{\Phi^+}_{23}$, the final state of the adiabatic evolution is given by $ U_2 \ket{\Phi^+}_{12} U_3^* \ket{\phi}_{3} $.

\begin{figure}[hbtp]
        \begin{center}
            \includegraphics[clip, width=0.7\columnwidth]{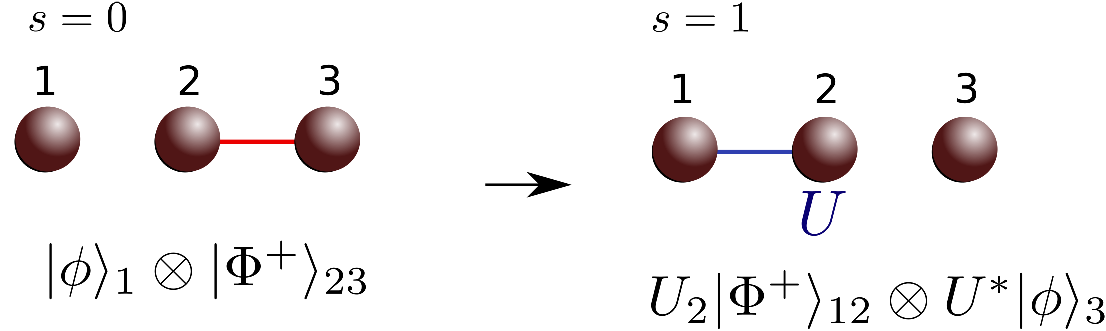}
        \end{center}
        \caption{
Implementation of complex conjugated unitary gate $U^*$.  The total Hamiltonian is given by $H^{U^*}_{\rm PAGT'} (\tau)=(1-s(\tau)) H_{\rm ini'}+ s(\tau) H_{\rm fin'}^{U^*} = (1-s(\tau)) H_{23} + s(\tau) H_{12}^U$.}
        \label{AGTfinal}
\end{figure}

To implement $U^\dagger$, we combine the previous two schemes and use the total Hamiltonian for adiabatic evolution as
\begin{equation}
	H^{U^\dagger}_{\rm PAGT'} (s(\tau))=(1-s(\tau)) H_{\rm ini'} + s(\tau) H_{\rm fin'}^{U^\dagger} \label{AGTdagger} 
\end{equation}
where $H_{\rm ini'}=H_{23}$ is unchanged from the case of $H^{U^*}_{\rm PAGT'} (\tau)$ but $H_{\rm fin'}^{U^\dagger} = H_{12}^{U^T}=H_{21}^{U}$.  Similarly to the case of implementing $U^*$, since
$H^{U^\dagger}_{\rm PAGT'} (\tau) = ( U_2^T  U_3^\dagger) H^{I}_{\rm PAGT} (\tau)  (U_2^T  U_3^\dagger )^\dagger$
holds due to Eq.~(\ref{twist3}), by choosing the initial state to be $\ket{\phi}_1 ( U_2^T  U_3^\dagger) \ket{\Phi^+}_{23} = \ket{\phi}_1 \ket{\Phi^+}_{23}$, the final state of adiabatic evolution is given by $ U_1 \ket{\Phi^+}_{12} U_3^\dagger \ket{\phi}_{3} $.

\begin{figure}[hbtp]
        \begin{center}
            \includegraphics[clip, width=0.7\columnwidth]{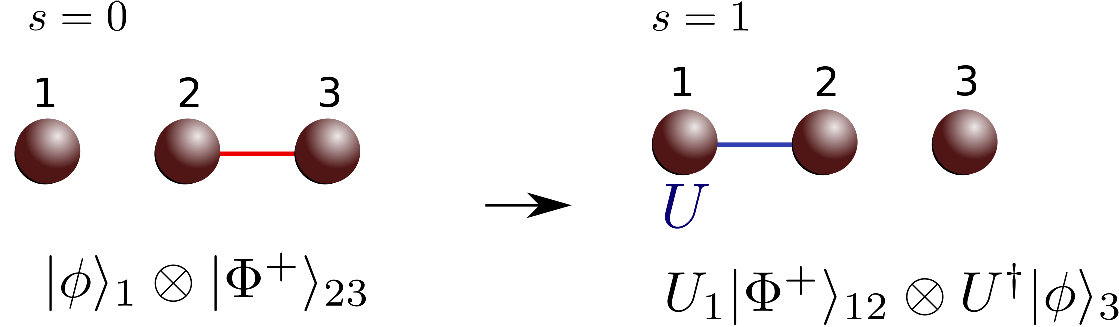}
        \end{center}
        \caption{
Implementation of daggered unitary gate $U^\dagger$.  The total Hamiltonian is given by $H^{U^\dagger}_{\rm PAGT'} (\tau)=(1-s(\tau)) H_{\rm ini'}+ s(\tau) H_{\rm fin'}^{U^\dagger} = (1-s(\tau)) H_{23} + s(\tau) H_{21}^U$.}
        \label{AGTfinal}
\end{figure}

We summarize the adiabatic gate teleportation schemes in TABLE.~\ref{tab-uni}.   An interesting property of the twisted Hamiltonian in PAGT is that it allows to implement $U^T$ and $U^*$ by using the twisted Heisenberg-type interaction  Hamiltonian given by $H_{\rm PAGT}^U$ for both cases and the ability to perform $U$ on a qubit for the initial state for case of implementing $U^T$.  This originates from the $-Y_i  Y_j$ term in the twisted Hamiltonian.  On the other hand, $U^\dagger$ can be implemented by the twisted interaction Hamiltonian without $-Y_i Y_j$,  $\tilde{H}_{\rm ini}^U$, which is shown in the Appendix B.

\begin{table*}[hbtp]
\begin{tabular}{|c|c|c|c|c|c|c|}
Scheme              & AGT($U$) & AGT'($U^\dagger$) &PAGT($U$)  & PAGT($U^{\rm T}$) &PAGT'($U^*$) & PAGT'($U^\dagger$)\\
\hline
Hamiltonian (notation)     &$H_{\rm AGT}^U(\tau)$&$H_{\rm AGT'}^{U^\dagger}(\tau)$& $H_{\rm PAGT}^U(\tau)$ & $H_{\rm PAGT}^{U^{\rm T}}(\tau) $& $H_{\rm PAGT'}^{U^*}(\tau) $  & $H_{\rm PAGT'}^{U^\dagger}(\tau)$ \\
Initial Hamiltonian      & $\tilde{H}_{23}^U$ & $\tilde{H}_{23}$ &${H}_{23}^U$ &${H}_{32}^U$&${H}_{23}$&${H}_{23}$\\
Final Hamiltonian       & $\tilde{H}_{12}$ & $\tilde{H}_{21}^U$&${H}_{12}$ &${H}_{12} $&${H}_{12}^U$ &${H}_{21}^U$\\
Heisenberg-type?   &No&No&Yes&Yes&Yes&Yes\\
Initial state                &$\ket{\phi}_{1}U_3\ket{\Phi^+}_{23}$&$\ket{\phi}_1 \ket{\Phi^+}_{23}$&$\ket{\phi}_{1}U_3\ket{\Phi^+}_{23}$ & $\ket{\phi}_{1}U_2\ket{\Phi^+}_{23}$ & $\ket{\phi}_{1}\ket{\Phi^+}_{23}$&$\ket{\phi}_{1}\ket{\Phi^+}_{23}$ \\
Final state             & $\ket{\Phi^+}_{12} U\ket{\phi}_{3}$& $U_1\ket{\Phi^+}_{12}U_1^\dagger \ket{\phi}_3$& $\ket{\Phi^+}_{12} U\ket{\phi}_{3}$ &$\ket{\Phi^+}_{12}U^{T}\ket{\phi}_{3}$& $U_2\ket{\Phi^+}_{12}U^{*}\ket{\phi}_{3}$ & $U_1\ket{\Phi^+}_{12} U^\dagger \ket{\phi}_{3}$ \\
\end{tabular}
\caption{Summary table of adiabatic gate teleportation schemes. The twisted interaction Hamiltonian used in AGT is defined by ${\tilde{H}_{ij}^U} : =-\omega  U_j ( X_i  X_j + Z_i  Z_j) U_j^\dagger$ and the twisted Heisenberg-type interaction Hamiltonian used in PAGT is defined by ${H_{ij}^U} : =-\omega  U_j ( X_i X_j - Y_i  Y_j + Z_i  Z_j) U_j^\dagger$.    We also define $\tilde{H}_{ij} :=\tilde{H}_{ij}^I$ and ${H}_{ij} :=H_{ij}^I$.  An arbitrary input state is denoted by $\ket{\phi}_1$ and $\ket{\Phi^+}=(\ket{00}+\ket{11})/\sqrt{2}$.    By using the twisted Heisenberg-type interaction  Hamiltonian $H_{ij}^U$, $U, U^T, U^*$ and $U^\dagger$  are implementable in the PAGT schemes.  By using the Hamiltonian $\tilde{H}_{ij}^U$, $U$ and $U^\dagger$ are implementable in the AGT scheme, but not $ U^T, U^*$ in general. 
}
\label{tab-uni}
\end{table*}

\section{Parallelization}\label{PAGT}

\subsection{Algorithm}\label{PAGTalgorithm}

We consider implementing {\it consecutive} $L$ single-qubit unitary gates $U^{(L)} \cdots U^{(2)}U^{(1)}$ on an input state $\ket{\phi}$ of a qubit system by using PAGT. If we try to implement $U^{(L)} \cdots U^{(2)}U^{(1)}$ in AGT, it can be achieved by iterating the AGT scheme $L$ times, namely we perform AGT to implement $U^{(1)}$ by using $H_{\rm AGT}^{U^{(1)}}$ and then perform another AGT to implement $U^{(2)}$ by using $H_{\rm AGT}^{U^{(2)}}$, and so on.   

\begin{figure}[htb]
	        \begin{center}
	            \includegraphics[clip, width=0.85\columnwidth]{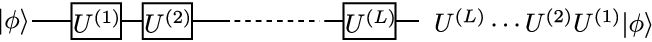}
	        \end{center}
	        \caption{A circuit implementation of consecutive $L$ single-qubit unitary gates $U^{(L)} \cdots U^{(2)}U^{(1)}$ applied on an input state $\ket{\phi}$ of a qubit system. }
	        \label{pAGTqc}
\end{figure}

We show that we can {\it parallelize} adiabatic gate teleportation by using the twisted Heisenberg-type interaction Hamiltonian including the $-Y_i  Y_j$ term ($H_{ij}^U$) for PAGT introduced in the previous section.  Our aim is to implement $U^{(L)}\cdots U^{(1)}$ on an input state $\ket{\phi}$ in a {\it single adiabatic evolution}.  

We use $2L+1$ qubit system $\ham_1 \otimes \ham_2 \otimes \cdots \otimes \ham_{2L+1}$.  We define the initial Hamiltonian to implement $U^{(L)}\cdots U^{(1)}$ as
\begin{equation}
	H_{\rm ini}^{U^{(L)}\cdots U^{(1)}} := \sum_{j=1}^L H_{2j ~ 2j+1}^{U^{(j)}},  \label{PAGThamini}
\end{equation}
and the final Hamiltonian as
\begin{equation}
	H_{{\rm fin}} := \sum_{j=1}^L H_{2j-1 ~ 2j}. \label{PAGThamfin}
\end{equation}

In the following, we prove that the adiabatic evolution of an initial state prepared in
\begin{equation}
	\ket{\Psi (0)}  = \ket{\phi}_1  U_{3}^{(1)} \ket{\Phi^+}_{23}  \cdots  U_{2L+1}^{(L)} \ket{\Phi^+}_{2L~2L+1}, \label{initialstatePAGT}
\end{equation}
under the total Hamiltonian given by
\begin{equation}
	H_{\rm PAGT}^{U^{(L)}\cdots U^{(1)}} (\tau):=(1-s(\tau))H^{U^{(L)}\cdots U^{(1)}}_{\rm ini}+s(\tau)H_{\rm fin}, \label{PAGTham}
\end{equation}
results in a final state
\begin{equation*}
	U^{(L)}_{2L+1}\cdots U^{(2)}_{2L+1} U^{(1)}_{2L+1} \ket{\phi}_{2L+1},
\end{equation*}
on the $(2L+1)$-th qubit, namely, we can achieve parallelized implementation of $U^{(L)} \cdots U^{(2)}U^{(1)}$ in PAGT.

\begin{figure}[H]
\begin{center}
    \includegraphics[clip, width=0.9\columnwidth]{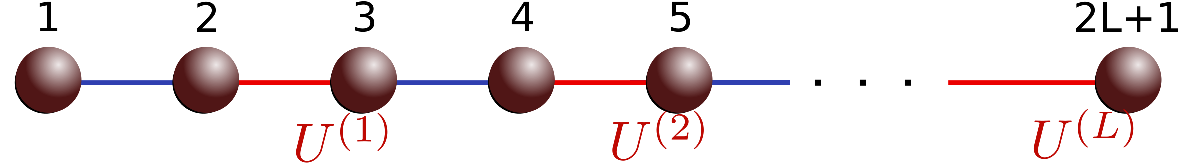}
\end{center}
\caption{(Color online) A schematic picture of parallelizable adiabatic gate teleportation (PAGT).  Lines between the $2j$-th and the $(2j+1)$-th qubits (red lines) represent the twisted Heisenberg-type interaction terms in the initial Hamiltonian $H^{U^{(L)}\cdots U^{(1)}}_{\rm ini}$.  Lines between the $(2j-1)$-th and the $2j$-th qubits (blue lines) represent the interaction terms in the final Hamiltonian $H_{\rm fin}$.}
\label{AGTparallel2}
\end{figure}

Before presenting the proof, we explain the intuition behind the PAGT scheme by comparing it with parallelized gate teleportation (PGT).  For arbitrary $\ket{\phi}_1$, the initial state $\ket{\Psi (0)}$ given by Eq.~(\ref{initialstatePAGT}) is a ground state of the initial Hamiltonian $H^{U^{(L)}\cdots U^{(1)}}_{\rm ini}$, since $ U_{2j+1}^{(j)} \ket{\Phi^+}_{2j~2j+1} \in \mathcal{H}_{2j} \otimes \mathcal{H}_{2j+1}$ is the ground state of the twisted Heisenberg-type interaction Hamiltonian $H_{2j~2j+1}^{U^{(j)}}$ for all $j$.  Similarly, the ground state of the final Hamiltonian $H_{\rm fin}$ is given by 
\begin{equation*}
 	\ket{\Psi (1)} = \ket{\Phi^+}_{12} \cdots \ket{\Phi^+}_{2L-1~2L} \ket{\phi '}_{2L+1} 
\end{equation*}
where $\ket{\phi '}_{2L+1}$ is an arbitrary state.  According to the adiabatic theorem, assuming the existence of the energy gap between the degenerated ground states and the first excited states throughout the time evolution,  the state $\ket{\Psi (0)}$ will be transformed to $\ket{\Psi (1)}$ if we drag $H^{U^{(L)}\cdots U^{(1)}}_{\rm ini}$ adiabatically to $H_{\rm fin}$.   Although the ground states are doubly degenerated, the adiabatic theorem given by \cite{Lidar}  is also applicable in the PAGT scheme, as we will show in the following that the logical operators are invariant under the adiabatic process similarly to the case of AGT presented in Section \ref{subsec:agt}.

In the PGT scheme, Bell measurements are performed on the $(2j-1)$-th and the $2j$-th qubits for $j=1, \cdots , L$ on the state given by $\ket{\Psi (0)}$. Considering the postselected case where all Bell measurement outcomes correspond to the projector $\ket{\Phi^+} \bra{\Phi^+}$, the output state is proportional to $\ket{\Psi (1)}$, where $\ket{\phi'}$ of the $(2L+1)$-th qubit is given by
\begin{equation*}
	\ket{\phi'} \propto ({U^{(L)}\cdots U^{(1)}})_{2L+1} \ket{\phi}_{2L+1}.
\end{equation*}
\begin{figure}[htb]
	        \begin{center}
	            \includegraphics[clip, width=0.7\columnwidth]{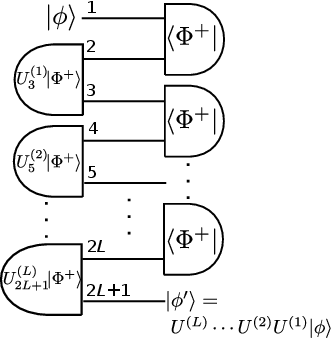} 
	        \end{center}
	        \caption{\label{PGTparallel} A circuit representations of the parallelized gate teleportation (PGT) scheme probabilistically implementing a sequence of unitary gates $U^{(L)}\cdots U^{(1)}$.  The left side of the circuit represents preparation of the initial state $\ket{\phi}_1U^{(1)}_3\ket{\Phi^+}_{23}\cdots U^{(L)}_{2L+1}\ket{\Phi^+}_{2L~2L+1}$ where $\ket{\phi}$ is an arbitrary input state.  The right side of the circuit with the symbol $\bra{\Phi^+}$ represents the postselection of Bell measurement outcome corresponding to the projector $\ket{\Phi^+} \bra{\Phi^+}$.  The postselected final state is given by $\ket{\Phi^+}_{12}\cdots \ket{\Phi^+}_{2L-1~2L} \ket{\phi '}$ where $\ket{\phi'}= (U^{(L)}\cdots U^{(1)})_{2L+1}\ket{\phi}_{2L+1}$.}
\end{figure}
The circuit representation of this scheme is shown in FIG.~\ref{PGTparallel}, where the symbol $\bra{\Phi^+}$ represents a postselection of the measurement outcome corresponding to the projector $\ket{\Phi^+} \bra{\Phi^+}$.   The key idea of PAGT is to {\it deterministically} simulate the input-output relation of the postselected circuit given in FIG.~\ref{PGTparallel} using an adiabatic evolution. 

Now we present the proof that the PAGT scheme works as expected.  We start by considering the simpler case of PAGT, where all gates $U^{(1)}, \cdots, U^{(L)}$ are the identity operators.  We call this special case as parallelizable adiabatic teleportation (PAT). The initial Hamiltonian is
\begin{equation}
	H_{\rm ini}^{I \cdots  I} := \sum_{j=1}^L H_{2j ~ 2j+1}, \label{PATini}
\end{equation}
and the final Hamiltonian is 
\begin{equation}
	H_{{\rm fin}} := \sum_{j=1}^L H_{2j-1 ~ 2j}. \label{PATfin}
\end{equation}
The total Hamiltonian for PAT is therefore given by
\begin{equation}
	H_{\rm PAT} (\tau):=(1-s(\tau))H^{I \cdots I}_{\rm ini}+s(\tau)H_{\rm fin}. \label{PATham}
\end{equation}

To check that the input state $\ket{\phi}_1 \in \ham_{1}$ is faithfully teleported to system $\ham_{2L+1}$ after the adiabatic evolution by $H_{\rm PAT}(\tau)$, we use techniques based on stabilizer formalism presented in the AGT scheme \ctr{30}.   We consider stabilizers of the initial Hamiltonian given by Eq.~\eqref{PATini} as
$\langle X_2X_3, Z_2Z_3, \cdots ,X_{2L}X_{2L+1},Z_{2L}Z_{2L+1} \rangle$ and the final Hamiltonian given by \Eqref{PATfin} as $\langle X_1X_2, Z_1Z_2, \cdots ,X_{2L-1}X_{2L},Z_{2L-1}Z_{2L}\rangle.$  The total system comprises $2L+1$ qubits and the stabilizer is given by $2L$ generators, which means the ground state is doubly-degenerate. We use these degrees of freedom to encode arbitrary qubit state by introducing two logical operators 
\begin{eqnarray*}
	L_X &:=& X_1X_2\cdots X_{2L+1}, \\
	L_Z &:=& Z_1Z_2 \cdots Z_{2L+1}. 
\end{eqnarray*}
Both $L_X$ and $L_Z$ commute with all stabilizer operators, that is,  $[L_k, H_{{\rm PAT}}(\tau)]=0$ for any $\tau\in [0,1]$ and $k=\{X,Z\}$.  Thus the subspace spanned by these logical operators is preserved throughout the evolution. The initial state is prepared in
\begin{equation*}
	\ket{\Xi (0)} = \ket{\phi}_{1} \ket{\Phi^+}_{23} \cdots \ket{\Phi^+}_{2L~2L+1},
\end{equation*}
which is the ground state of $H_{\rm PAT}(0) = H^{I \cdots I}_{\rm ini}$.  At $\tau=0$, we have
\begin{eqnarray*}
 	L_X \ket{\Xi (0)} &=& X_1 (X_2X_3)\cdots (X_{2L}X_{2L+1}) \ket{\Xi (0)} \\
 	&=& X_1 \ket{\Xi (0)} , \\
 	L_Z \ket{\Xi (0)} &=& Z_1 \ket{\Xi (0)},
\end{eqnarray*}
namely, $L_X$ and $L_Z$ are represented by $X_1$ and $Z_1$, respectively at $\tau=0$.  

The ground state of the total Hamiltonian at the final time is given by $ \ket{\Xi (1)} = \ket{\Phi^+}_{12} \cdots \ket{\Phi^+}_{2L-1~2L} \ket{\phi '}_{2L+1}$, where $\ket{\phi'}_{2L+1}$ is an arbitrary state in $\ham_{2L+1}$.  Since 
\begin{eqnarray*}
	L_X  \ket{\Xi (1)} &=& X_{2L+1} \ket{\Xi (1)}, \\
	L_Z  \ket{\Xi (1)} &=& Z_{2L+1} \ket{\Xi (1)}, 
\end{eqnarray*}
and the logical operators at the initial and final times have to correspond to each other, the following conditions
\begin{eqnarray*}
	\bra{\phi} X \ket{\phi} &=&  \bra{\phi'} X \ket{\phi'}, \\
	\bra{\phi} Z \ket{\phi} &=&  \bra{\phi'} Z \ket{\phi'}, 
\end{eqnarray*} 
must hold. Therefore we obtain $\ket{\phi'} = \ket{\phi}$, that is, teleportation of a state $\ket{\phi}$ from the first qubit to the (2L+1)-th qubit is faithfully achieved. 

Next we consider the general cases of the PAGT scheme where the total Hamiltonian $H_{{\rm PAGT}}^{U^{(L)}\cdots  U^{(1)}}(\tau)$ is given by Eq.~\eqref{PAGTham}.   For simplicity, we recursively define a sequence of $l$ unitary gates $W_i^{(l)}=U_i^{(l)}\cdots  U_i^{(1)}$ on system $\mathcal{H}_i$ by
\begin{equation}
	W^{(l)}_i :=U_i^{(l)} W_i^{(l-1)}\quad\text{and}\quad W_i^{(0)} :=I_i, \label{defW}
\end{equation}
for $l=1,\cdots,L$.
Using the property of the twisted Heisenberg-type interaction  Hamiltonian presented in \Eqref{twist3},  $H_{{\rm PAGT}}^{U^{(L)}\cdots  U^{(1)}}$ can be written in terms of $H_{{\rm PAT}}$ as
\begin{equation}
	H_{{\rm PAGT}}^{U^{(L)}\cdots  U^{(1)}}(\tau)=\tilde{V}^{{W^{(L)}}} H_{{\rm PAT}}(\tau) \left(\tilde{V}^{W^{(L)}}\right)^\dagger, \label{PAGThami}
\end{equation}
 where $\tilde{V}^{{W^{(L)}}}$ is a unitary operator on system $\ham_1 \otimes \ham_2 \otimes \cdots \otimes \ham_{2L+1}$ defined by
\begin{equation*}
	\tilde{V}^{{W^{(L)}}}:= \left ( \bigotimes_{l=1}^{L}  W_{2l-1}^{(l-1)} W_{2l}^{(l-1)*}  \right) \otimes W_{2L+1}^{(L)},
\end{equation*}
since 
\begin{align}
	&\tilde{V}^{{W^{(L)}}}  \left \{ \sum_{j=1}^L H_{2j ~ 2j+1} \right \} (\tilde{V}^{W^{(L)}})^\dagger \nonumber \\
	&= \sum_{j=1}^L W_{2j}^{(j-1)*} W_{2j+1}^{(j)} (H_{2j ~ 2j+1}) (W_{2j}^{(j-1)*} W_{2j+1}^{(j)} )^\dagger \nonumber \\
	&=  \sum_{j=1}^L U_{2j+1} ^{j} H_{2j ~ 2j+1} (U_{2j+1} ^{j})^\dagger = \sum_{j=1}^L H_{2j ~ 2j+1}^{U^{(j)}}.
\end{align}
and 
\begin{align}
	&\tilde{V}^{{W^{(L)}}}  \left \{ \sum_{j=1}^L H_{2j-1 ~ 2j} \right \} (\tilde{V}^{W^{(L)}})^\dagger \nonumber \\
	&= \sum_{j=1}^L W_{2j-1}^{(j)} W_{2j}^{(j)*} (H_{2j-1 ~ 2j}) (W_{2j-1}^{(j)} W_{2j}^{(j)*})^\dagger \nonumber \\
	&=  \sum_{j=1}^L H_{2j-1 ~ 2j}.
\end{align}
The initial states of $H_{{\rm PAGT}}^{U_L^{(L)} \cdots U_1^{(1)}}(\tau)$ and  $H_{{\rm PAT}}$ are also related via $\tilde{V}^{W^{(L)}}$ as $\ket{\Phi (0)} =\tilde{V}^{W^{(L)}}\ket{\Xi (0)}$. Therefore if there is no energy level crossing between the ground states and the first excited states, the adiabatic evolution by the total Hamiltonian given by $H_{{\rm PAGT}}^{U_L^{(L)} \cdots U_1^{(1)}}(\tau)$ transform an initial state $\ket{\Phi (0)}=\ket{\phi}_1 U_3^{(1)}\ket{\Phi^+}_{23} \cdots U_{2L+1}^{(L)}\ket{\Phi^+}_{2L~2L+1}$ to the desired final state given by $\ket{\Phi (1)} = \ket{\Phi^+}_{12}\cdots\ket{\Phi^+}_{2L-1~2L} W_{2L+1}^{(L)} \ket{\phi}$ for arbitrary $\ket{\phi}_1$. 

\subsection{Energy gap and Time of Computation}

A sequence of $L$ unitary gates can be performed by either iterating the AGT scheme $L$ times, or applying the single-step PAGT scheme. We compare the AGT and PAGT schemes by considering the scaling of the total time $T_L$ required for adiabatic evolution with respect to the number of unitary gates $L$. Since the AGT scheme utilizes a fixed system size with a fixed minimum energy gap to perform each of the unitary gates, the total time increases linearly with $L$. The scaling for the PAGT scheme can be obtained by using \Eqref{totaltime2} by considering the scaling of each term. We start by considering the minimum energy gap of the total Hamiltonian.

Let us define $\Delta E_L (\tau)$ as the energy gap of an $L$-gate PAGT scheme, and $\mathcal{G}_L := \min_ {\tau\in[0,1]} \Delta E_L (\tau)$ to be the minimum energy gap.   The minimum energy gap of AGT is equivalent to the minimum energy gap of PAGT for $L=1$,  $\Delta E_1(\tau)$ represents the energy gap of the AGT scheme using the total Hamiltonian given by Eq.~\eqref{ATham}.
We analyze the scaling behavior of the energy gap $\Delta E_L(\tau)$.

Since the energy eigenvalues do not change under unitary conjugation, the energy gap of the total Hamiltonian $H_{\rm PAGT}^{W^{(L)}}(\tau)$ given by \Eqref{PAGTham} implementing PAGT and the total Hamiltonian $H_{\rm PAT}(\tau)$ given by \Eqref{PATham} are same.
Moreover, by  conjugating $H_{\rm PAT}(\tau)$ with $\prod_{j=1}^L Y_{2j}$, that is applying the Pauli $Y$ operators on qubits with even number indices, we obtain a new Hamiltonian $H_{{\rm s-chain}}(\tau)$, defined by
\begin{eqnarray}
 H_{{\rm s-chain}}(\tau)&:=& \left(\prod_{j=1}^L Y_{2j}\right) \cdot H_{{\rm PAT}}(\tau) \cdot \left(\prod_{j=1}^L Y_{2j}\right)^\dagger \nonumber\\
 &=&\omega\sum_{j=1}^L \left[\left(1-s(\tau)\right) {\bm S_{2j}}\cdot{\bm S_{2j+1}}\right. \nonumber \\
 &+&\left.s(\tau) {\bm S_{2j-1}}\cdot{\bm S_{2j}}\right]. 
\label{antiferro}
\end{eqnarray}
Exploiting the invariance of the energy spectrum under unitary conjugation, it suffices to analyze $H_{{\rm s-chain}}(\tau)$ which is an alternating bond 1-D anti-ferromagnetic spin chain Hamiltonian. For $s=1/2$, the Hamiltonian corresponds to an open boundary 1-D XXX Heisenberg spin chain model with size $2L+1$.

The analysis of the energy gap of $H^{{\rm s-chain}}(\tau)$ can be simplified by considering a symmetry of the Hamiltonian and noting the existence of invariant subspaces throughout the evolution (see also \ctr{54}).
The total spin, defined by $J_z:=\frac{1}{2}\sum_{i=1}^{2L+1}Z_i$, commutes with the Hamiltonian in Eq.~(\ref{antiferro}), $[J_z,H^{{\rm s-chain}}(\tau)]=0$. We define subspace denoted by $\mathcal{S}_k $ as
\begin{equation}
 \mathcal{S}_k := \text{span} ~ \{ \ket{s} ~ | ~ J_z \ket{s} = k \ket{s} \},
\end{equation}
where $k=-(2L+1)/2, \cdots , -1/2, 1/2,\cdots, (2L+1)/2$.
All $2L+2$ subspaces are invariant under the evolution, i.e. $H_{{\rm s-chain}}(\tau)\mathcal{S}_k \subseteq \mathcal{S}_k$.  The two degenerate ground states $\ket{0}_L$ and $\ket{1}_L$ are in the subspace $\mathcal{S}_{1/2}$ and $\mathcal{S}_{-1/2}$, respectively.   The logical state $\ket{0}_L$ is transformed only to another state inside the subspace $\mathcal{S}_{1/2}$ during the evolution generated by $H_{{\rm s-chain}}(\tau)$. 
Therefore it is sufficient to only consider the energy spectrum inside $\mathcal{S}_{\frac{1}{2}}$. 
The same logic applies to the logical state $\ket{1}_L$ in $\mathcal{S}_{-1/2}$. 

The total spin represents an important symmetry of the Hamiltonian $H_{{\rm s-chain}}(\tau)$ which remains unchanged after flipping all spins, $Z_i \rightarrow -Z_i,~1 \leq i \leq 2L+1 $.
The properties of the energy gap concerning $\mathcal{S}_k$ are the same as the properties of $\mathcal{S}_{-k}$, thus it suffices to consider only the energy gap of the ground states and the first excited states in the subspace $\mathcal{S}_{\frac{1}{2}}$. 
In the rest of the paper, $\Delta E_L(\tau)$ denotes the energy gap in this subspace.

\begin{figure}[]
 \begin{center}
  \includegraphics[clip, width=0.7\columnwidth,angle=-90]{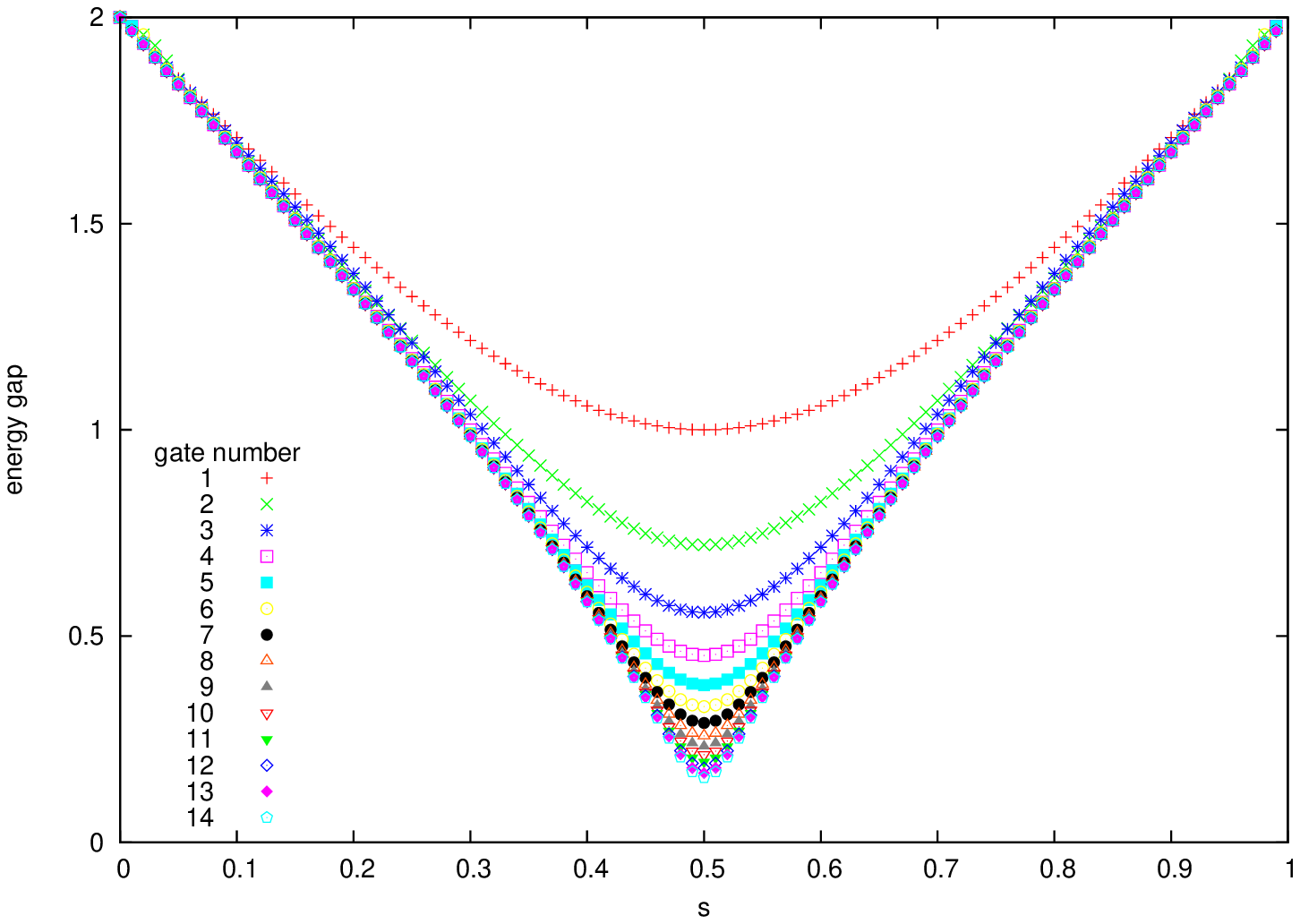}
 \end{center}
 \caption{
(Color online) Energy gap $\Delta E_L(\tau)$ with respect to the interpolation function $s(\tau)$. The energy gap between the ground states and the first excited states in \Eqref{antiferro} are shown. We set $\omega=0.5$ and plot $\Delta E_L(\tau)$ for each $s=0.01, 0.02, \cdots , 1.00$. We change color for each $L$ (gate number) from 1 to 14, which appear in the order from top to bottom.}
 \label{fig:energygap}
\end{figure}

\begin{figure}[]
	\begin{center}
		\includegraphics[clip, width=0.7\columnwidth, angle=-90]{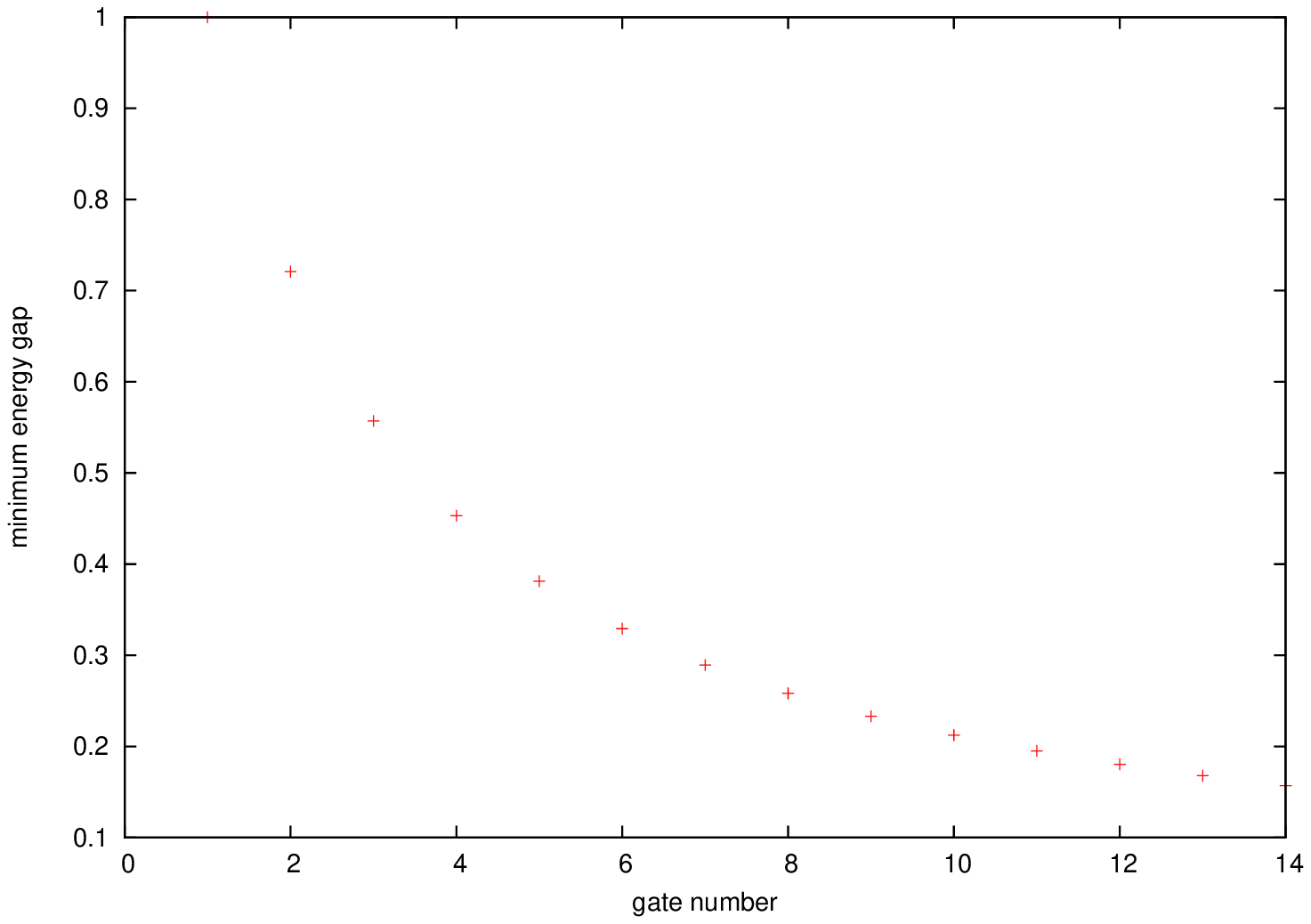}
 	\end{center}
 	\caption{Minimum energy gap $\mathcal{G}_L$ with respect to $L$ (gate number). The data are obtained from FIG.~\ref{fig:energygap} at $s=1/2$.}
 \label{fig:minE}
\end{figure}

We have numerically studied this energy gap, by making use of the TITPack ver.2 which is a FORTRAN library for diagonalizing quantum spin Hamiltonians developed by Nishimori \cite{TITPack}.   We have calculated $\Delta E_L(\tau)$ by setting $\omega=0.5$, with $s= 0.01, 0.02, \cdots , 1.00$ for each $L$ between $1$ to $14$.  The result is displayed in \Figref{fig:energygap}.  The minimum energy gap $\mathcal{G}_L$ is obtained at $s=1/2$. As mentioned above, the Hamiltonian at this point becomes a 1-D XXX Heisenberg model whose minimum spectral gap scales as $O(1/L)$. This scaling is also supported by our numerical results in \Figref{fig:minE}.

The next term in Eq.~(\ref{totaltime2}) that we consider is the largest ``curvature" of the time-dependent Hamiltonian $\xi$. In Appendix \ref{app:ham_norm}, we construct a specific interpolation function $s(\tau)$ that does not scale with the size of the system $L$ and show that
\begin{equation}
 \xi=\sup_{\tau}|\partial_{\tau}s(\tau)|\cdot\|H_{{\rm fin}}-H_{{\rm ini}}\| =O(L). \label{max-spectrum}
\end{equation}

Lastly, the height of the analyticity domain $\gamma$ in Eq.~\eqref{totaltime2} is independent of the system size $L$. Therefore we conclude for the total time $T_L=O(L^5)$. This shows that parallelization of AGT can be achieved with a polynomial overhead in terms of the number $L$ of consecutive unitary gates. The main advantage of PAGT lies in its ability to manipulate temporal order by mapping the time-aligned temporal order of gates to the spatial order of interactions in the final Hamiltonian as described in the next subsection.

\subsection{Gate order manipulation}
In PAGT, twisted interaction Hamiltonians $H_{\rm PAGT}^{U^{(1)}}, \cdots, H_{\rm PAGT}^{U^{(L)}}$ are simultaneously applied.  The implemented operations $W^{(L)}=U^{(L)}\cdots U^{(1)}$ are ordered even if the time-dependent strength of each twisted Heisenberg-type interaction Hamiltonian determined by the parameter $s(\tau)$ is the same for all interaction Hamiltonians during the evolution. 

We clarify that the order of applied gates is determined only by the form of the final Hamiltonian.  Let us demonstrate this on a particular example when $L=2$.    In the PAGT scheme using the total Hamiltonian $H_{\rm PAGT}^{W^{(2)}}$, $U^{(2)} U^{(1)}$ is applied to the input state. 
If the final Hamiltonian is modified to
\begin{align}
	{H'}_{{\rm fin}} &:= H_{{14}} + H_{{25}} \ntg
	&= -\omega [( X_{1}X_{4}-Y_{1} Y_{4}+Z_{1} Z_{4} ) \ntg
	&~~~~~ +(X_{2}X_{5}-Y_{2} Y_{5}+Z_{2} Z_{5})], \label{order2}
\end{align}
instead of $H_{\rm fin}$ given by \Eqref{PAGThamfin}, the final state of adiabatic evolution is given by $U^{(1)} U^{(2)}\ket{\phi}_3$.

One of the characteristics of the PAGT scheme is to encode information about the unitary gates ({\it what} unitary gates to perform) and their order ({\it which} order to perform) {\it separately} to the initial and final Hamiltonians, respectively.   This property leads to the ability to coherently control the order of gates by adding a control-qubit in the adiabatic scheme. 

\section{Controlled parallelized adiabatic gate teleportation (C-PAGT)} \label{CPAGT}

In this section, we show that several differently ordered PAGT evolutions with the same initial Hamiltonian can be ``superposed'' by adding a control-qubit. We refer to this scheme as {\it controlled parallelized adiabatic gate teleportation} (C-PAGT). It utilizes the properties similar to the PAGT scheme where the roles of the initial and final Hamiltonian are separated, namely, the initial Hamiltonian determines which gates to apply and the final Hamiltonian determines the order of gates to be applied.  The C-PAGT scheme implements the $n$-SWITCH gate described in Section \ref{sec:qs}. The temporal order of gates can be manipulated by the state of the control-qubit.  In the following, we consider $n=2$ given by $U_{\rm QS}^{F,G}$ (the $2$-SWITCH gate) introduced in Section \ref{sec:qs} for simplicity since generalization to the case of $n>2$ is straightforward.

We consider a 6-qubit system $\ham_{C} \otimes \ham_{1} \otimes \ham_{2} \otimes \ham_{3} \otimes \ham_{4} \otimes \ham_{5}$, where $\ham_C$ denotes the system of the control-qubit. The input state is prepared in $\ket{\phi}_{C 1}=\ket{0}_C\ket{\phi_0}_1+\ket{1}_C\ket{\phi_1}_1 \in \ham_{C}\otimes \ham_{1}$, where $\ket{\phi_0}$ and $\ket{\phi_1}$ are arbitrary, not necessarily orthogonal, single-qubit states. The initial Hamiltonian is given by
\begin{equation}
	H_{\rm ini}^{FG}= H_{23}^F+H_{45}^G, \label{QSini}
\end{equation}
where $H_{23}^F$ and $H_{45}^G$ are twisted Heisenberg-type interaction Hamiltonians defined by Eq.~(\ref{gateham2}). Note that $H_{\rm ini}^{F G}$ acts trivially on $\ham_C$ and $\ham_1$ and thus the initial Hamiltonian has four-fold degenerate ground states. We choose an initial state (a ground state $H_{\rm ini}^{FG}$) as
\begin{equation}
	\ket{\Psi(0)} := \ket{\phi}_{C1} F_3\ket{\Phi^+}_{23} G_5\ket{\Phi^+}_{45}. \label{QSinistate}
\end{equation}

We design the final Hamiltonian to be able to control the order of the unitary gates $F$ and $G$ depending on the state of the control-qubit by introducing interaction between the control-qubit and the rest of the system,
\begin{equation}
	H_{{\rm fin}}^{{\rm C-PAGT}} = \kb{0}_C \otimes {H}_{{\rm fin}}^{(0)}+\kb{1}_C \otimes {H}_{{\rm fin}}^{(1)}, \label{controlledHam} 
\end{equation}
where
\begin{eqnarray*}
	{H}_{{\rm fin}}^{(0)} &:=& H_{12} + H_{34} , \\
	{H}_{{\rm fin}}^{(1)} &:=& H_{14} + H_{25}.
\end{eqnarray*}
Note that this final Hamiltonian does not depend on the gates we want to perform. The final Hamiltonian for C-PAGT in Eq.~(\ref{controlledHam}) can be written as
\begin{equation}
	H_{{\rm fin}}^{{\rm C-PAGT}} := \frac{1}{2} ({H}_{{\rm fin}}^{(0)}+{H}_{{\rm fin}}^{(1)} + Z_C \otimes {H}_{{\rm fin}}^{(0)} -Z_C \otimes {H}_{{\rm fin}}^{(1)}), \label{controlledHam22} 
\end{equation}
and includes 3-body interactions $Z_C \otimes {H}_{{\rm fin}}^{(0)}$ and $Z_C \otimes {H}_{{\rm fin}}^{(1)}$.
The total Hamiltonian for adiabatic evolution for C-PAGT is given by
\begin{equation}
	H_{{\rm C-PAGT}}^{F,G}(\tau) := (1-s(\tau))H^{FG}_{\rm ini} +s(\tau) H_{{\rm fin}}^{{\rm C-PAGT}} . \label{AGTQS}
\end{equation}

\begin{figure}[hbt]
	        \begin{center}
	            \includegraphics[clip, width=0.7\columnwidth]{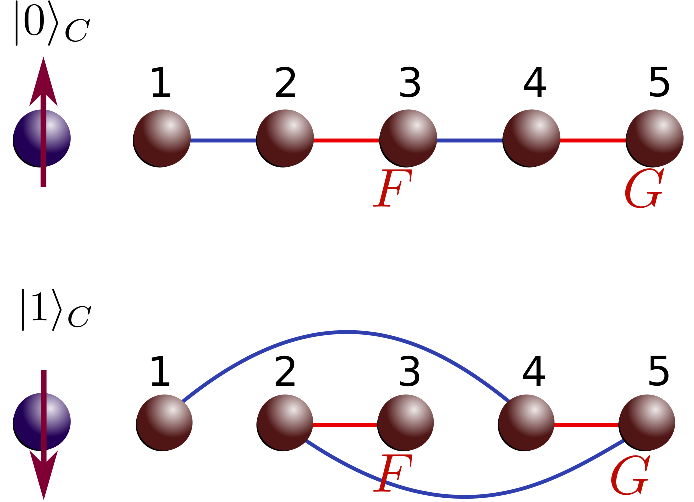} 
	        \end{center}
	        \caption{ (Color online) Schematic representations of conditional transformations corresponding to the two parts of the final Hamiltonian of C-PAGT, $H_{{\rm fin}}^{{\rm C-PAGT}}$.  The upper figure corresponds to the term $ \kb{0}_C \otimes {H}_{{\rm fin}}^{(0)}$ and the lower figure corresponds to the term} $ \kb{1}_C \otimes {H}_{{\rm fin}}^{(1)}$.  The blue lines represent interaction Hamiltonians in the final Hamiltonians.   The red lines represent twisted Heisenberg-type interaction  Hamiltonians in the initial Hamiltonian $H^{FG}_{\rm ini} $.  
\label{figAGTQS}
\end{figure}

Now we prove that the adiabatic evolution generated by $H_{{\rm C-PAGT}}^{F,G}(\tau)$ implements the 2-SWITCH gate $U_{\rm QS}^{F,G}$.   In the simplest case of $F=G=I$, the corresponding Hamiltonian is denoted as $H_{{\rm C-PAT}}(\tau)$.  We call this case controlled parallelized adiabatic teleportation (C-PAT).  Similarly to the case of PAGT, the total Hamiltonian in Eq.~(\ref{AGTQS}) can be written as a unitary conjugation of the C-PAT Hamiltonian
\begin{equation*}
	H_{{\rm C-PAGT}}^{F,G} (\tau) = \tilde{V}^{F,G}  H_{{\rm C-PAT}}(\tau) (\tilde{V}^{F,G})^\dagger, 
\end{equation*}
where
\begin{align}
	&\tilde{V}^{F,G} \nonumber\\
	&=\kb{0}_C \otimes F_3F_4^* (G F)_5 + \kb{1}_C \otimes G_2^* (FG)_3 G_5. \label{Vunitary}
\end{align}
Therefore proving that C-PAT works immediately shows that C-PAGT works as well.

We begin by showing that using adiabatic evolution generated by $H_{{\rm C-PAT}}(\tau)$, we can superpose adiabatic teleportation of the initial state $\ket{\phi}_{C1}=\ket{0}_C\ket{\phi_0}_1+\ket{1}_C\ket{\phi_1}_1$ to the final state $\ket{0}_C\ket{\phi_0}_5+\ket{1}_C\ket{\phi_1}_3$. We may try to define the conjugated logical $X$ operator as $L_X^{\rm Control}:={\rm CNOT}_{C5} (\prod_{i=1}^5X_i) {\rm CNOT}_{C5}$, where the ${\rm CNOT}_{Ci}$ acts on system $\mathcal{H}_C \otimes \mathcal{H}_i$. However $L_X^{\rm Control}$ does not commute with $H_{{\rm C-PAT}}(\tau)$ and so the stabilizer approach used in previous sections does not apply any more and we have to consider the dynamics explicitly. $H_{{\rm C-PAT}}(\tau)$ can be rewritten as
\begin{align}
	H_{{\rm C-PAT}}(\tau) &=\kb{0}_C \otimes \left[ (1-s(\tau))H^{II}_{{\rm ini} ~\backslash C}  +s(\tau) H_{{\rm fin}}^{(0)} \right] \nonumber\\
	&+\kb{1}_C \otimes \left[ (1-s(\tau))H^{II}_{{\rm ini} ~\backslash C}  +s(\tau) H_{{\rm fin}}^{(1)} \right], \label{qsat}
\end{align}
where $H^{II}_{{\rm ini} ~\backslash C}$ denotes the Hamiltonian obtained after tracing out the control-qubit system $\ham_C$ of $H^{II}_{{\rm ini}}$.  The Hamiltonian in Eq.~(\ref{qsat}) can be written in a succinct form as
\begin{equation}
	H_{{\rm C-PAT}}(\tau) = \sum_{j=0}^1\kb{j}_C \otimes H^{(j)} (\tau), \label{BDform}
\end{equation}
where $H^{(j)} (\tau) := (1-s(\tau)) H^{II}_{{\rm ini} ~\backslash C} + s(\tau) H_{{\rm fin}}^{(j)}$.

Due to the block diagonal form of Eq.~(\ref{BDform}), the time evolution operator of $H_{{\rm C-PAT}}(\tau)$ can be written in terms of $H^{(j)} (\tau)$ as  
\begin{equation*}
	\mathcal{T} e^{-i\int_0^{\tau} dt' H_{{\rm C-PAT}}(t')} = \sum_{j} \kb{j}_C\otimes \mathcal{T} e^{-i\int_0^{\tau} dt' H^{(j)} (t')}, 
\end{equation*}
where $\mathcal{T}$ is the time-ordering operator.  For an initial state given by
\begin{equation}
	\ket{\Xi (0)}=\sum_j \ket{j}_C \ket{\Xi_j}_{12345}, \label{Xiinitial}
\end{equation}
for $j=0,1$, and where $\ket{\Xi_0}_{12345}:= \ket{\phi_0}_1 \ket{\Phi^+}_{23} \ket{\Phi^+}_{45}$ and $\ket{\Xi_1}_{12345}:= \ket{\phi_1}_1 \ket{\Phi^+}_{23} \ket{\Phi^+}_{45}$. The adiabatic evolution by $H_{{\rm C-PAT}}(\tau)$ transfers the initial state to  
\begin{eqnarray}
	\ket{\Xi (\tau)} &:=& \mathcal{T} e^{-i\int_0^{\tau} dt' H^{{\rm C-PAT}}(t') }\ket{\Xi (0)} \nonumber\\
	&=& \sum_j \ket{j}_C  \mathcal{T} e^{-i\int_0^{\tau} dt' H^{(j)} (t') } \ket{\Xi_j}_{12345} \nonumber\\
	&=& \sum_j \ket{j}_C  \ket{\Xi_j (\tau)}_{12345}.
\end{eqnarray}

The Hamiltonian $H^{(j)} (\tau)$ has the same form as the PAGT Hamiltonian in \Eqref{PAGTham} with $L=2$ and $\ket{\Xi_j}_{12345}$ is the ground state of $H^{(j)} (\tau)$. Therefore the adiabatic evolution of the state $\ket{\Xi_j (\tau)}$ is guaranteed to be the same as in the case of the PAGT scheme. Thus the adiabatic evolution by $H^{(j)}(\tau)$ is guaranteed to transform
\begin{eqnarray}
	\ket{\Xi_0}_{12345} &\rightarrow & \ket{\Phi^+}_{12}\ket{\Phi^+}_{34}\ket{\phi_0}_5 , \nonumber\\
	\ket{\Xi_1}_{12345} &\rightarrow & \ket{\Phi^+}_{14}\ket{\Phi^+}_{25}\ket{\phi_1}_3. 
\end{eqnarray}
To show successful implementation of $U_{\rm QS}^{F,G}$, we have to show that the relative phase between these states remains 0.

In this case, the corresponding $H^{(0)}$ and $H^{(1)}$ are given by
\begin{align}
	H^{(0)} (\tau)&=(1-s(\tau) ) \left( H_{23} + H_{45} \right) 
+ s (\tau) \left( H_{12} + H_{43} \right),  \label{h0} \\
	H^{(1)} (\tau) &= (1-s(\tau) ) \left( H_{23} + H_{45} \right) 
+ s (\tau) \left( H_{14} + H_{25} \right), 
\end{align}
By relabelling the Hilbert spaces $\ham_{1} \otimes \ham_{2} \otimes \ham_{3} \otimes \ham_{4} \otimes \ham_{5}$ on which $H^{(1)} (t)$ acts according to $2 \leftrightarrow 4$ and $3\leftrightarrow 5$, we obtain a new Hamiltonian $H^{' (1)}$ such that 
\begin{eqnarray*}
	H^{' (1)} (\tau) &=& (1-s(\tau) ) \left( H_{23} + H_{45} \right) 
+ s (\tau) \left( H_{12} + H_{43} \right). 
\end{eqnarray*}
Comparing this expression with \Eqref{h0}, we see the relative phase between $\ket{\Xi_0}$ and $\ket{\Xi_1}$ is guaranteed to be 0.  Therefore  
the adiabatic evolution generated by this Hamiltonian transforms the initial state $\ket{\Xi (0)}$ to
\begin{eqnarray*}
	\ket{\Xi^\prime (1)} &=& \ket{0}_C \ket{\Phi^+}_{12}\ket{\Phi^+}_{34}\ket{\phi_0}_5 \\
	&+& \ket{1}_C \ket{\Phi^+}_{14}\ket{\Phi^+}_{25}\ket{\phi_1}_3.
\end{eqnarray*}

Now we consider the case of $F \neq I$ and $G \neq I$ using the relationship $H_{\rm C-PAGT}^{F,G} (\tau) = \tilde{V}^{F,G} H_{{\rm C-PAT}}(\tau) (\tilde{V}^{F,G})^\dagger$ where $\tilde{V}$ is given by Eq.~(\ref{Vunitary}).  The initial states in \Eqref{QSinistate} and \Eqref{Xiinitial} are related, namely $\ket{\Psi (0)} = \tilde{V}^{F,G} \ket{\Xi (0)}$. So the final state at the end of the adiabatic evolution is given by
\begin{eqnarray*}
	\ket{\Psi (1)} &=& \tilde{V}^{F,G} \ket{\Xi^\prime  (1)} \\
	&=& \ket{0}_C \ket{\Phi^+}_{12}\ket{\Phi^+}_{34} GF \ket{\phi_0}_5 \\
	&+& \ket{1}_C \ket{\Phi^+}_{14}\ket{\Phi^+}_{25} FG \ket{\phi_1}_3.
\end{eqnarray*}
We still need to ensure that for each conditional case, the output qubit is in $\mathcal{H}_5$. This can be done by applying a controlled SWAP gate non-adiabatically. CSWAP$_{ij}$ is a three-qubit unitary operation on $\ham_C \otimes \ham_i \otimes \ham_j$ defined by
\begin{equation*}
	\text{CSWAP}_{ij} := \ket{0} \bra{0}_C \otimes I_{ij}+ \ket{1} \bra{1}_C \otimes \text{SWAP}_{ij}, 
\end{equation*}
where SWAP$_{ij}$ is a swap gate on $\ham_i \otimes \ham_j$ that maps $\ket{a}_i \ket{b}_j \rightarrow \ket{b}_i \ket{a}_j$, for any $\ket{a}$ and $\ket{b}$.
By applying CSWAP$_{24}$ and CSWAP$_{13}$ on $\ket{\Psi(1)}$, the final state is transformed to
\begin{equation*}
    \ket{\Phi^+}_{12}\ket{\Phi^+}_{34} \left(\ket{0}_C GF \ket{\phi_0}_5 + \ket{1}_C FG\ket{\phi_1}_5\right),
\end{equation*}
which is the state obtained by performing $U_{\rm QS}^{F,G}$.

Before concluding the proof, we still need to check that there is no energy level crossing between the degenerate ground states and the degenerate first excited states.  
We denote the eigenvalues and the eigenstates of $H^{(j)} (\tau)$ as $E^{(j)}_i(\tau)$ and $\ket{\Psi^{(j)}_i (\tau)}$ for $j=0,1,$ and $0 \le i \le 2^{5}-1$. Namely, 
\begin{equation}
	H^{(j)} (\tau) \ket{\Psi^{(j)}_i (\tau)}=E^{(j)}_i (\tau) \ket{\Psi^{(j)}_i (\tau)}, \label{eigenparallel}
\end{equation}
where $i~(j)$ represents the number of degrees of freedom of the logical target-qubit (control-qubit), respectively. 

The eigenvalues and the eigenstates of the controlled version of $H^{{\rm C-PAGT}}(\tau)$ are given by 
\begin{equation*}
	H_{{\rm C-PAGT}}^{FG} (\tau) \ket{j}_C \ket {\Psi^{(j)}_i (\tau)}= E^{(j)}_i (\tau) \ket{\Psi^{(j)}_i (\tau)}.
\end{equation*}
Therefore the eigenvalues of the total Hamiltonian of adiabatic evolution remain unchanged from the corresponding PAGT scheme described by $H^{(j)} (\tau)$. The computational time to successfully implement the C-PAGT scheme for $L$ consecutive gates is the same with the corresponding PAGT scheme. 

We have shown that transformation implemented by the quantum switch can be simulated using the C-PAGT scheme. 
It is achieved by a single use of two interaction twisted Hamiltonians $H_F$ and $H_G$.  In the quantum circuit model, it is straightforward to count the number of oracles to evaluate the computational resource of the circuit, but in adiabatic computation, the ``number of calls'' is not well-defined.  Therefore, a fair comparison between the computational resources of the two models is not straightforward.  We note that this result can be easily generalized to any number of consecutive single-qubit gates
\begin{equation*}
	\sum_{x=0}^{L!} \ket{x} \bra{x}_C \otimes U^{(\sigma_x(L))} \cdots U^{(\sigma_x(2))} U^{(\sigma_x(1))},
\end{equation*}
where $\sigma_x (i)$ denotes the permutation function $\{1,2, \cdots , L\} \mapsto \{1,2, \cdots , L\}$.  However in this case, the description of the final Hamiltonian grows super-exponentially.

\section{Implementability of time-reversal Hamiltonian dynamics and parallelizability}
\label{TRHdynamics}

In this section, we analyze parallelizability of a sequence of unitary gates in the adiabatic gate teleportation schemes in terms of implementability of transposed unitary gates.  We show that the ability to implement transposed unitary gates is closely related to implementability of dynamics generated by time-reversal Hamiltonian in the adiabatic schemes.

In classical mechanics, the time-reversal transformation reverses the time parameter $t$ to $-t$. When acting on a Hamiltonian in classical mechanics, terms depending on the momentum are time-reversed while terms depending on the position remain invariant.  In quantum mechanics, the time-reversal transformation $\Theta$ is defined by a combination of applying a complex conjugate operation in terms of a fixed basis denoted by $\ast$ and a unitary operator $V$, namely  $\Theta = \ast V $, where  $V$ satisfies $V^\ast V=e^{i\phi}I$, where $\phi$ denotes a global phase factor. Note that the complex conjugate operation $\ast$ depends on the basis and there is a freedom of choice of the basis for time-reversal operations.  This condition for the unitary operator $V$ originates from the requirement of the self-inverse property of $\Theta$ up to a global phase factor, $\Theta^2 = e^{i\phi}I$.  

We denote by $A^R$ the observable obtained by the time-reversal transformation of observable $A$. Then $A^{\rm R}$ is represented by  $A^{\rm R} = \Theta A \Theta^\dagger$, where $\Theta^\dagger = V^\dagger \ast$.  As an example, let us consider the case when $\Theta= \ast$ (i.e. $V=I$), where the complex conjugate is taken in the position basis.  For a free particle system described by its position observable $x$ and momentum observable $p$, we have $x^{\rm R} = x$ and $p^{\rm R} = - p$.   For a spin-1/2 system,  the time-reversal operator is given as $\Theta = \ast Y$.   In this case, for any spin-observable represented by ${\bm S}= (X{\bm e}_x + Y{\bm e}_y  + Z {\bm e}_z )/2$ where ${\bm e}_x, {\bm e}_y, {\bm e}_z$ are unit vectors for the $x$-axis, $y$-axis and $z$-axis, respectively, we have ${\bm S}^{\rm R} = -{\bm S}$. 

The dynamics generated by a time-independent Hamiltonian $H$ is represented by a unitary evolution operator $U=e^{- i H t}$ (using units where $\hbar=1$).  The dynamics generated by the time-reversal Hamiltonian $H^{\rm R}=\Theta H \Theta^\dagger$ (which we call {\it TRH dynamics} for short) is unitarily equivalent to $U^{\rm T}$ where ${\rm T}$ denotes a transposition in terms of a pre-fixed basis.   This can be easily checked as
\begin{eqnarray}
 	e^{-iH^{\rm R}t}  &=& V^\ast e^{-i  H^\ast  t} V^{\rm T} \nonumber\\
 	&=& V^\ast \left(e^{i H t}\right)^{\ast} V^{\rm T} \nonumber\\
 	&=& V^\ast  U^{\rm T} V^{\rm T}.
 \label{TRHdynamics1}
\end{eqnarray}
Note that in general the TRH dynamics is different from $U^R = V^\ast U^\ast V^{\rm T}$, which is sometimes called as  the ``time reversed'' dynamics.

In the following, we consider how to implement the TRH dynamics $e^{-iH^{\rm R}t}$ given by Eq.~(\ref{TRHdynamics1}) using only interaction Hamiltonians implementing $U=e^{- i H t}$.  The unitary part $V$ is a fixed unitary operation and is independent of the choice of $U$.   However, there is no known way to universally and deterministically map $U$ to $U^{\rm T}$ in the quantum circuit model in general. In contrast, we have shown in Section \ref{power} that implementation of $U^{\rm T}$ is possible by using the twisted Heisenberg-type interaction Hamiltonian $H^U_{ij}$ and changing the spatial order of interactions. We compare the ability to implement $U^{\rm T}$ in both AGT and PAGT, and study the relationship between TRH dynamics and parallelizability in PAGT.

We first show that the AGT scheme cannot implement $U^{\rm T}$ in general by using the twisted interaction Hamiltonian implementing $U$ in the same way that PAGT does in Section \ref{power}.  Assume that $U^{\rm T}$ is implementable in the AGT scheme by just using twisted interaction terms in the form of $\tilde{H}_{ji}^U$, where $j>i$.  The corresponding total Hamiltonian of AGT $H_{\rm AGT}^{U^{\rm T}}$ is given by
\begin{align*}
	H_{\rm AGT}^{U^{\rm T}}(\tau) = (1-s(\tau))\tilde{H}_{32}^{U} + s(\tau)\tilde{H}_{12}.
\end{align*}
The two logical operators of AGT are defined by 
\begin{align*}
	L_X^{U^{\rm T}}&:=X_1X_2 X_3^{{U^{\rm T}}}, \\
	L_Z^{U^{\rm T}}&:=Z_1Z_2 Z_3^{{U^{\rm T}}},
\end{align*}
where we use the notation that a conjugation of an operator $A$ by a unitary operator $U$ is represented by $A^{U}:=UA U^\dagger$.  

For successful implementation of $U^{\rm T}$ using $H_{\rm AGT}^{U^{\rm T}}$, 
the logic operators and the total Hamiltonian must satisfy the commutation relations given by
\begin{equation}
	[L_X^{U^{\rm T}},H_{\rm AGT}^{U^{\rm T}} (\tau)] = [L_Z^{U^{\rm T}},H_{\rm AGT}^{U^{\rm T}} (\tau)]=0, \label{condition4L}
\end{equation}
for all $\tau \in [0,1]$.
We have
\begin{eqnarray*}
	[L_X^{{U^{\rm T}}},H_{\rm AGT}^{U^{\rm T}}(\tau)] &=& [ X_1 X_2 X_3^{U^{\rm T}}, (1-s(\tau))\tilde{H}_{32}^{U} + s(\tau)\tilde{H}_{12}]  \\
	&=& (1-s(\tau)) X_1 [X_2 X_3^{U^{\rm T}},\tilde{H}^{U}_{32}].
\end{eqnarray*}
The commutation relation can be further transformed to
\begin{eqnarray*}
	[X_2 X_3^{U^{\rm T}},\tilde{H}^{U}_{32}] &=& U_2 [U_2^\dagger X_2 U_2 X_3^{U^{\rm T}} , \tilde{H}_{23} ]U_2^\dagger \\
	&=& U_2[X_2^{U^\dagger} X_3^{U^{\rm T}},\tilde{H}_{23},  ]U_2^\dagger.
\end{eqnarray*} 
Thus $[X_2^{U^\dagger} X_3^{U^{\rm T}},\tilde{H}_{23} ]=0$ must be satisfied for Eq.~\eqref{condition4L} to hold. A similar expression can be obtained for logical operator ${L_Z^{U^{\rm T}}}$. Therefore Eq.~\eqref{condition4L} is satisfied if and only if 
\begin{equation}
	[X_2^{U^\dagger} X_3^{U^{\rm T}}, \tilde{H}_{23}]=[Z_2^{U^\dagger} Z_3^{U^{\rm T}},\tilde{H}_{23}]=0 \label{comm}
\end{equation}
holds, where $\tilde{H}_{23}$ is the interaction Hamiltonian given by \Eqref{AGTinteration}.

The condition given by \Eqref{comm} is not satisfied by a general $U$. In Appendix \ref{time-inverse}, we present the condition on $U$ so that Eq.~(\ref{comm}) is satisfied and show that any $U$ satisfying Eq.~(\ref{comm}) also satisfies
\begin{equation}
	[X_2^{U}X_3^{U^\ast},\tilde{H}_{23}]=[Z_2^{U}Z_3^{U^\ast},\tilde{H}_{23}]=0. \label{comm2}
\end{equation}

Finally, we clarify when a parallel implementation of a sequence of $L$ unitary gates  ${U^{(L)} \cdots U^{(1)}}$ is possible with the AGT Hamiltonian on system $\ham_1 \otimes \ham_2 \otimes \cdots \otimes \ham_{2L+1}$ defined by 
\begin{eqnarray}
	H_{\rm AGT}^{{U^{(L)} \cdots U^{(1)}}}(\tau) &:=& (1-s(\tau)) \sum_{j=1}^L \tilde{H}_{2j~ 2j+1}^{U^{(j)}} \nonumber\\
	&+& s(\tau) \sum_{j=1}^L \tilde{H}_{2j-1~2j} \label{AGTparallel}
\end{eqnarray}
in a {\it single} step without using the twisted Heisenberg-type interaction Hamiltonian.  The difference between the AGT Hamiltonian and the PAGT Hamiltonian is just the form of the two-body interactions $\tilde{H}_{ij}^U$ and ${H}_{ij}^U$, namely the PAGT Hamiltonian includes additional $- Y_i Y_j$ terms.

The ability of parallelization in PAGT introduced in Sec.~\ref{PAGTalgorithm} can be described by commutation relations between the total Hamiltonian and the logical operators in the stabilizer formalism.  Using the notation of $W^{(L)}$ introduced in Eq.~(\ref{defW}),  the two logical operators are represented by
\begin{align*}
	L^{U^{(L)}\dots U^{(1)}}_X &:=  \tilde{V}^{{W^{(L)}}} (X_1 \cdots X_{2L+1}) (\tilde{V}^{{W^{(L)}}})^\dagger \\
	&= \left( \bigotimes_{l=1}^{L} X_{2l-1}^{W^{(l-1)}} X_{2l}^{W^{(l-1)\ast}}  \right) \otimes X_{2L+1}^{W^{(L)}}, \\
 	L^{U^{(L)}\dots U^{(1)}}_Z &:=   \tilde{V}^{{W^{(L)}}} (Z_1 \cdots Z_{2L+1}) (\tilde{V}^{{W^{(L)}}})^\dagger \\ 
	&= \left( \bigotimes_{l=1}^{L} Z_{2l-1}^{W^{(l-1)}} Z_{2l}^{W^{(l-1)\ast} } \right) \otimes Z_{2L+1}^{W^{(L)}}.
\end{align*}
Therefore, the commutation relations must satisfy
\begin{align}
	[L^{U^{(L)}\dots U^{(1)}}_X, H_{\rm AGT}^{{U^{(L)} \cdots U^{(1)}}} (\tau)]&=0,\label{commcondini} \\
	[L^{U^{(L)}\dots U^{(1)}}_Z, H_{\rm AGT}^{{U^{(L)} \cdots U^{(1)}}} (\tau)]&=0, \label{commcondfin}
\end{align}
for all $\tau \in [0,1]$ for successful implementation of the sequence of $L$ unitary gates in a single step.  
Since the commutators of each term in Eq.~(\ref{AGTparallel}) and $L^{U^{(L)}\dots U^{(1)}}_X$ are linearly independent to each other, the above equations are satisfied if and only if, 
\begin{align}
	[X_{j+1}^{W^{(\lfloor j/2 \rfloor)}}X_{j+2}^{W^{(\lfloor j/2 \rfloor)\ast}},\tilde{H}_{j+1 ~ j+2}] = 0,& \label{equicommini} \\
	[Z_{j+1}^{W^{(\lfloor j/2 \rfloor)}}Z_{j+2}^{W^{(\lfloor j/2 \rfloor)\ast}},\tilde{H}_{j+1 ~ j+2}] = 0,&  \label{equicommfin}
\end{align}
for any $0 \leq j \leq 2L-1$  (see Appendix D for the details).

The commutation relations given by Eqs.~(\ref{equicommini}) and (\ref{equicommfin}) 
are in the same form of Eq.~(\ref{comm2}). Therefore a sequence of unitary gates $U^{(1)},\dots, U^{(L)}$ are implementable in AGT if $(W^{(l)})^{\rm T}$, for all $l$, is also implementable. In the PAGT scheme, similar commutation relations obtained by replacing $\tilde{H}_{j+1~j+2}$ with $H_{j+1~j+2}$ in Eq.~(\ref{AGTparallel}) are automatically satisfied for any $U$ since the additional $- Y_{j+1} Y_{j+2}$ term in $H_{j+1~j+2}$ makes the final Hamiltonian satisfy $H^{U}_{j~j+1}=H^{U^{\rm T}}_{j+1~j}$.   By using this condition, for any unitary $U$, $H_{j~j+1}$ satisfies  
\begin{eqnarray}
  [X_j^{U}X_{j+1}^{U^\ast}, H_{j~j+1}] &=&  U_{j+1}^\ast [X_j^{U} X_{j+1},H_{j~j+1}^{U^{\rm T}} ]U_{j+1}^{\rm T} \nonumber \\
  &=& U_{j+1}^\ast [X_j^{U} X_{j+1},H_{j+1~j}^{U} ]U_{j+1}^{\rm T} \nonumber \\
  &=& U_j U_{j+1}^\ast [X_j X_{j+1},H_{j~j+1} ] U_j^\dagger U_{j+1}^{\rm T} \nonumber \\
  &=& 0.
\end{eqnarray}
Thus $U^{\rm T}$  is always implementable in PAGT model and the ability of the TRH dynamics can be considered to enable the PAGT scheme to be for any set of $\{ U^{(i)}\}$. That is, deterministic simulation of a postselected event of parallelized gate teleportation is achieved in adiabatic implementation using the PAGT scheme.

\section{Conclusion}
\label{conclusion}

In this paper, we have investigated a method to manipulate the temporal order of a sequence of unitary gates in the setting of adiabatic quantum computation. Our method is based on the adiabatic gate teleportation (AGT) scheme proposed by Bacon and Flammia \ctr{30}.   We introduce {\it twisted Heisenberg-type interaction  Hamiltonians} to implement a unitary  gate in an adiabatic manner and show that the transpose, complex conjugate, and adjoint of the unitary gate are also implementable by just changing the spatial order of interactions and/or exchanging the initial and final Hamiltonians.

Using the twisted Heisenberg-type interaction  Hamiltonians, we construct a parallelizable adiabatic gate teleportation (PAGT) scheme, where consecutive gate operations are implemented in a single adiabatic step.  An important feature of the PAGT scheme is that information about the unitary gate and information about the temporal order of gates are separately encoded in the initial Hamiltonian and the final Hamiltonian respectively. We can choose the total unitary gate by changing the twisted Heisenberg-type interaction Hamiltonian in the initial Hamiltonian, and we can choose the temporal order of these unitary gates by changing the spatial order of interactions of the final Hamiltonian. This property enables us to construct the controlled parallelized adiabatic teleportation (C-PAGT) scheme that uses a control-qubit to manipulate the temporal order of the unitary gates.  

On the other hand, PAGT offers no advantage in terms of the computational time required to implement $L$ consecutive gate operations compared to the AGT scheme since the energy gap narrows by parallelization.   The total time scales as $O(L^5)$ with the number of unitary gates $L$, as opposed to $O(L)$ in the AGT case where the scheme is simply iterated $L$ times.

We investigate why the twisted Heisenberg-type interaction Hamiltonian allows PAGT.  We show that this interaction Hamiltonian has an ability to perform a transposed unitary gate by just modifying the spatial ordering of the final Hamiltonian implementing a unitary gate in adiabatic gate teleportation. The dynamics generated by the time-reversed Hamiltonian represented by the transposed unitary gate enables deterministic simulation of a postselected event of parallelized gate teleportation in adiabatic implementation.

%%%%%%%%%%%%%%%%%%%%%%%%%%%%%ACKNOWLEDGMENTS
\section*{Acknowledgments}
We would like to acknowledge Elham Kashefi for introducing adiabatic gate teleportation to us .
We also would like to thank Shunsuke Furukawa for helpful discussions on numerical methods on spin systems and Nathan Wiebe for his suggestions about modern adiabatic theorems.. This work is supported by the Project for Developing Innovation Systems of MEXT, Japan, the Global COE Program of MEXT Japan, and JSPS KAKENHI (Grant No.~23540463, No.~23240001, No.~26330006, and No.~15H01677).  We also gratefully acknowledge to the ELC project (Grant-in-Aid for Scientific Research on Innovative Areas MEXT KAKENHI (Grant No.~24106009)) for encouraging the research presented in this paper. MH also acknowledges support from Singapore's National Research Foundation and Ministry of Education. This material is based on research funded by the Singapore National Research Foundation under NRF Award NRF-NRFF2013-01. 
SN also acknowledge the Foundational Questions Institute through``The fundamental principles of information dynamics'' (FQXi-RFP3-1325).''

\appendix

\section{Implementation of a daggered unitary gate in AGT}
We show that a daggered unitary gate ($U^\dagger$) is implementable in the AGT scheme using the initial Hamiltonian given by 
\begin{equation*}
	H_{{\rm ini'}}= -\omega(X_2 X_3 + Z_2 Z_3) = \tilde{H}_{23},
\end{equation*}
and the final Hamiltonian given by
\begin{equation*}
	H_{{\rm fin'}}^{{U^\dagger}}= - \omega U_2(X_1 X_2 + Z_1 Z_2)U_2^\dagger  =  \tilde{H}_{21}^U.
\end{equation*}
The total Hamiltonian is given by
\begin{eqnarray*}
 	H_{\rm AGT'}^{{U^\dagger}}(\tau) &:=& (1-s(\tau))   H_{{\rm ini'}}+s(\tau)  H_{{\rm fin'}}^{{\rm U^\dagger}} \\
	&=&U_1 H^I_{{\rm AGT}}(\tau) U_1^\dagger,
\end{eqnarray*}
where $H^I_{{\rm AGT}}(\tau)$ is given by \Eqref{ATham}.   Using this total Hamiltonian, an initial state prepared in $U_1 \ket{\phi}_1 \ket{\Phi^+}_{23}$ is transformed to a final state given by $U_1\ket{\Phi^+}_{12}\ket{\phi}_3$ under the adiabatic evolution.  By setting $\ket{\phi'}:= U\ket{\phi}$, we can understand the input state $\ket{\phi'}_1$ at $\tau=0$ is transformed to $U ^\dagger\ket{\phi'}_3$ at $\tau=1$.   Therefore, the $-Y_i Y_j$ term included in the total Hamiltonian of the PAGT scheme $H_{\rm PAGT'}^{{U^\dagger}}(\tau) $ is not necessary for implementing $U^\dagger$. 

\section{Interpolation function $s(\tau)$ and linear scaling of the largest curvature $\xi$}
\label{app:ham_norm}
We explicitly construct the interpolation function $s(\tau)$ and show that the largest curvature of the Hamiltonian scales linearly with the system size $L$. As explained in Section \ref{adiabatic theorem}, in order to apply the adiabatic theorem, the interpolation function for a Hamiltonian $H(\tau)=(1-s(\tau))H_{\text{ini}}+s(\tau)H_{\text{fin}}$ must satisfy the following properties,
\begin{enumerate}
	\item If $\tau\in[0,1]$, then $s(\tau)\in\mathbb{R}$.
	\item $s(0)=0$ and $s(1)=1$.
	\item $s(z)$ is holomorphic in $z\in\{z|\text{dist}(z,[0,1])\leq\gamma\}$ for some $\gamma>0$.
	\item $\partial_{\tau}^n s(0)=\partial_{\tau}^n s(1)=0$, for $1\leq n\leq N$, where $\tau=\text{Re}[z]$.
\end{enumerate}
For $z,w\in\mathbb{C}$, satisfying $\text{Re}[z],\text{Re}[w]>0$, the \textit{incomplete beta function} $B_{\tau}(z,w)$ is defined as
\begin{equation*}
	B_{\tau}(z,w)=\int_0^\tau dt~t^{z-1}(1-t)^{1-w}.
\end{equation*}
The usual \textit{beta function} is obtained for $\tau=1$, that is $B(z,w)=B_1(z,w)$. We can define the \textit{regularized incomplete beta function} $I_{\tau}(z,w)$ as
\begin{equation*}
	I_{\tau}(z,w)=\frac{B_{\tau}(z,w)}{B(z,w)}
\end{equation*}
For $x,y\in\mathbb{R}$ and $x,y>0$, we have that $B(x,y)>0$, since $\tau^{x-1}(1-\tau)^{y-1}>0$ when $\tau\in[0,1]$. Therefore $I_{\tau}(x,y)$ is a well-defined real-valued function on $\tau\in[0,1]$. We refer to this function as $f_{x,y}(\tau)=I_{\tau}(x,y)$. This function has the property that
\begin{equation*}
	f_{x,y}(0)=0\qquad\text{and}\qquad f_{x,y}(1)=1.
\end{equation*}
For any $x,y>N$,
\begin{equation*}
	\partial_{\tau}^n f_{x,y}(0)=\partial_{\tau}^n f_{x,y}(1)=0,
\end{equation*}
for $1\leq n \leq N$. Expanding $f_{x,y}(\tau)$ as a polynomial of real variable $\tau$ and replacing $\tau$ with $z\in\mathbb{C}$, we see that $f_{x,y}$ is a complex function which is holomorphic in the entire complex plane. This shows that $f_{x,y}(\tau)$, for $x,y>N$, satisfies all conditions for Theorem 1 to apply. We obtain the desired interpolation function by setting $x,y=N+1$, that is
\begin{equation*}
	s(\tau)=f_{N+1,N+1}(\tau).
\end{equation*}

Next we focus on the largest curvature of the time-dependent Hamiltonian,
\begin{eqnarray}
	\xi &=& \sup_{\tau}\left\|\partial_{\tau} H(\tau)\right\| \nonumber\\
	&=& \sup_{\tau}\left|\partial_{\tau}s(\tau)\right|\cdot\left\|H_{\text{fin}}-H_{\text{ini}}\right\|. \label{eq:curvature}
\end{eqnarray}
In particular, we want to determine the scaling of Eq.~\eqref{eq:curvature} with respect to $L$.
The derivative of the interpolation function is independent of $L$ and scales only with the controllable parameter $N$. Therefore the only term that is $L$-dependent is the operator norm of the difference between the final and initial Hamiltonians.

To show that $\|H_{\text{fin}}-H_{\text{ini}}\|$ scales linearly with $L$, we calculate a lower and upper bound for the operator norm. The lower bound can be obtained by considering the definition of the operator norm, $\left\| A \right\| \geq \left\|A\ket{v_0} \right\|$, for some arbitrary unit vector $\ket{v_0}$. By choosing $\ket{v_0}=\ket{00110011 \cdots 00110}$, we see that
\begin{equation*}
	\left\|H_{\text{fin}} - H_{\text{ini}} \right\| \ge \bra{v_0} H_{{\rm fin}} - H_{{\rm ini}} \ket{v_0} = \omega L,
\end{equation*}
which shows that the lower bound scales as $O(L)$. The upper bound is derived by using the triangle inequality
\begin{eqnarray*}
 \left\|H_{{\rm fin}} - H_{{\rm ini}} \right\| &=& \left\|\omega \sum _{k=1}^{2L} (-1)^k {\bm S}_k \cdot {\bm S}_{k+1} \right\| \\
 &\leq & \omega \sum _{k=1}^{2L} \left\| {\bm S}_k \cdot {\bm S}_{k+1} \right\| = 6 \omega L
\end{eqnarray*}
Therefore we can conclude that $\|H_{\text{fin}}-H_{\text{ini}}\|=O(L)$ and also that the largest curvature $\xi=O(L)$.

\section{Condition of implementing TRH dynamics in AGT}\label{time-inverse}

A general unitary operation on a Hilbert space $\mathcal{H}={\mathbb C}^2$ can be represented by
\begin{align*}
 	&U(x,y,z) = \cos x \cdot I  \\
 	& - i \sin x \left[  \cos y\cdot Y  
 	+ \sin y \left(  \cos z\cdot X + \sin z\cdot Z \right)   \right],
\end{align*}
where $x, y, z \in [-\pi, \pi ) $.  Using this notation, the commutation relations required for the AGT Hamiltonian and logic operators are given by 
\begin{align}
	[X^{U^\dagger}_2 X_3^{U^{\rm T}}, {H_{\rm AGT}^{I}}] =& a(x,y,z) \cdot (I_2 X_3 - X_2I_3) \nonumber  \\
  	&- b(x,y,z)\cdot(I_2Z_3 - Z_2 I_3)  \label{comx}
 \end{align}
and
 \begin{align}
[Z^{U^\dagger}_2 Z_3^{U^{\rm T}}, {H_{\rm AGT}^{I}}] =& c(x,y,z)\cdot (I_2 X_3 - X_2I_3) \nonumber  \\
  &- d(x,y,z)\cdot (I_2Z_3 - Z_2 I_3), \label{comz}
\end{align}
where 
\begin{eqnarray*}
 	a(x,y,z) &=&- 8 i \sin^2 x  \sin y \cdot f(x,y,z) p(x,y,z), \\
 	b(x,y,z) &=& - 8 i \sin x  \sin y \cdot f(x,y,z) u(x,y,z), \\
 	c(x,y,z) &=& -4 i \sin x  \sin y \cdot g(x,y,z) v(x,y,z), \\
 	d(x,y,z) &=&-16 i \sin^2 x \sin y \cdot g(x,y,z) q(x,y,z), 
\end{eqnarray*}
and 
\begin{eqnarray*}
 	f(x,y,z) &=&  \sin x \cos y  \cos z - \cos x  \sin z, \\
	g(x,y,z) &=&  \sin x \cos y \sin z + \cos x  \cos z, \\
	 p(x,y,z) &=& \sin x \sin^2 y \sin z \cos z+\cos x \cos y, \\
 	q(x,y,z) &=& \sin x \sin^2 y \sin z \cos z -\cos x \cos y, \\
 	u(x,y,z) &=& \cos^2 x - \sin^2 x ( \cos^2 y -  \sin^2 y \cos 2z), \\
 	v(x,y,z) &=& \cos^2 x - \sin^2 x ( \cos^2 y +  \sin^2 y \cos 2z).
\end{eqnarray*}

As discussed in the main text, the TRH dynamics of $U$ is implementable if and only if the commutators in Eq.~(\ref{comx}) and Eq.~(\ref{comz}) both vanish.
This condition can be written as $a=b=c=d=0$ and is equivalent to either one or to a logical disjunction of the following expressions which are exclusive to each other.
\begin{enumerate}
 	\item $\sin x = 0$.
 	\item $\sin x \neq 0$, $\sin y = 0$.
 	\item $\sin x , \sin y \neq 0$, $f = g = 0$.
 	\item $\sin x, \sin y, g \neq 0$, $f = q = v= 0$. 
 	\item $\sin x , \sin y , f \neq 0$, $g = p = u= 0$.
 	\item $\sin x, \sin y, f, g \neq 0$, $ p = q = u = v= 0$.
\end{enumerate}
The mutual exclusiveness of the above conditions is not necessary in nature. However it does avoid confusion in what follows. 

Conditions 3 to 6 can be simplified as we show now.
\paragraph*{Condition 3.} Since $f$ and $g$ satisfy the following equation 
\begin{equation*}
 	\cos z \cdot g - \sin z \cdot f = \cos x,
\end{equation*}
then $\cos x = 0$. By substituting $\cos x =0$ to $f$ and $g$,  we also obtain $\cos y =0$.

\paragraph*{Condition 4.} $f$ and $q$ satisfies the following equation 
\begin{equation}
 	\cos y \cdot f - \sin z \cdot q = \sin x \cos z (\cos^2 y - \sin^2 y \sin^2 z). \label{for4}
\end{equation}
Since the above equation is zero, one (or both) of the following cases is true. 
\subparagraph{4-a. $\cos z = 0$. } By substituting this condition to $f$, then we obtain $\cos x \sin z = 0$.  Since $\sin z =  \pm 1$ (coming from $\cos z  = 0 $), $\cos x = 0$.  Substituting $\cos z = \cos x = 0$ to $v$, we obtain 
\begin{equation*}
 	v = - ( \cos^2 y - \sin^2 y  ) = -\cos 2 y.
\end{equation*}
Since $v=0$. the following conditions are satisfied, 
\begin{equation*}
 	\cos x = \cos 2y = \cos z = 0. 
\end{equation*}
Substituting the above condition to $g$, we obtain
\begin{equation*}
 	g = \pm \cos y.
\end{equation*}
Using the condition that $g\neq0$, we must have that $\cos y = \pm 1/\sqrt{2}$.  Then we obtain
\begin{equation}
	\sin x= \pm 1, \quad \sin z = \pm 1, \quad \cos y = \pm \frac{1}{\sqrt{2}}. \label{con4}
\end{equation}

\subparagraph{4-b. $\cos^2 y = \sin^2 y \sin^2 z$.} Note that the following equality is satisfied, 
\begin{align*}
 \cos^2 y + &\sin^2 y \cos 2z \\ 
 = & (\cos^2 y - \sin^2 y \sin^2 z) + \sin^2 y \cos^2 z.
\end{align*}
Then $v$ is simplified to 
\begin{equation*}
 	v = \cos^2 x - \sin^2 x \sin^2 y \cos^2 z.
\end{equation*}
Because $v=0$, we obtain the following two equations,
\begin{eqnarray}
 	\cos^2 x &=& \sin^2 x \sin^2 y \cos^2 z, \nonumber\\
 	\cos^2 y &=& \sin^2 y \sin^2 z. \label{sq4}
\end{eqnarray}
By the normalization, the above equations imply
\begin{align*}
 	\cos^2 x + \sin^2 x & ( \cos^2 y + \sin^2 y (\cos^2 z + \sin^2 z))  \\
 	&= 2 \cos^2 x + 2 \sin^2 x \cos^2 y \\
 	&= 2 - 2\sin^2 x \sin^2 y \\
 	&=1 . 
\end{align*}
Then 
\begin{equation*}
 	\sin^2 x \sin^2 y = \frac{1}{2}.
\end{equation*}
Combining the above equation with Eq.~(\ref{sq4}), we have
\begin{eqnarray}
 	\sin x \sin y &=& \pm\frac{1}{\sqrt{2}}, \label{sqcon41}  \\
 	\cos x &=& \pm \frac{1}{\sqrt{2}} \cos z,  \label{sqcon42} \\
 	\cos y &=& \pm \frac{1}{\sqrt{2}}. \label{sqcon43}
\end{eqnarray}
Using Eq.~(\ref{sqcon43}), we obtain $\sin y = \pm1/\sqrt{2}$.  Substituting this condition to Eq.~(\ref{sqcon41}), $\sin x=\pm 1$.  Then $\cos x = 0$.  Combining this and Eq.~(\ref{sqcon42}), $\cos z= 0$ is satisfied.
This leads to same expressions as in Eq.~(\ref{con4}).

Let us check the inverse. Using Eq.~(\ref{con4}), we can quickly confirm that indeed $f=q=v=0$ and that $g=\pm 1/\sqrt{2}$.

\paragraph*{Condition 5.} $g$ and $p$ satisfy the following equation,
\begin{equation}
 	\cos y \cdot g - \cos z \cdot p = \sin x \sin z (\cos^2 y - \sin^2 y \cos^2 z). \label{for5}
\end{equation}
Since the above equation is zero, one (or both) of the following cases is true. 
\subparagraph{5-a. $\sin z = 0$.}  By substituting this condition to $g$, we obtain $\cos x \cos z = 0$.  Since $\cos z = \pm 1$ (using $\sin z  = 0 $), $\cos x = 0$.  Substituting $\sin z = \cos x = 0$ to $u$, we obtain 
\begin{equation*}
	 u = - ( \cos^2 y - \sin^2 y  ) = - \cos 2y.
\end{equation*}
Since $u=0$, the following expression is satisfied, 
\begin{equation}
 	\cos x = \cos 2y = \sin z = 0. \label{con5a}
\end{equation}
Substituting the above condition to $f$, we obtain
\begin{equation*}
 	g = \pm \cos y.
\end{equation*}
Since $f$ is non-zero, $\cos y = \pm 1/\sqrt{2}$ must be satisfied.  We then obtain
\begin{equation}
 	\sin x= \pm 1, \quad \cos z = \pm 1, \quad \cos y = \pm \frac{1}{\sqrt{2}}. \label{con5}
\end{equation}

\subparagraph{5-b. $\cos^2 y = \sin^2 y \cos^2 z$.}  Note that the following equality is satisfied, 
\begin{align*}
 	\cos^2 y - &\sin^2 y \cos 2z \\ 
 	= & (\cos^2 y - \sin^2 y \cos^2 z) + \sin^2 y \sin^2 z.
\end{align*}
Then $u$ can be simplified to 
\begin{equation*}
 	u = \cos^2 x - \sin^2 x \sin^2 y \sin^2 z.
\end{equation*}
Because $u=0$,  we obtain the following two equations,
\begin{equation}
	\cos^2 x = \sin^2 x \sin^2 y \sin^2 z,~ \cos^2 y = \sin^2 y \cos^2 z. \label{sq5}
\end{equation}
By the normalization, the above equations imply
\begin{align*}
 	\cos^2 x + \sin^2 x & ( \cos^2 y + \sin^2 y (\cos^2 z + \sin^2 z))  \\
 	&= 2 \cos^2 x + 2 \sin^2 x \cos^2 y \\
 	&= 2 - 2\sin^2 x \sin^2 y \\
 	&=1 . 
\end{align*}
Using this, we can write that
\begin{equation*}
 	\sin^2 x \sin^2 y = \frac{1}{2}.
\end{equation*}
Combining the above equation with Eq.~(\ref{sq4}), we have
\begin{eqnarray}
 	\sin x \sin y &=& \pm\frac{1}{\sqrt{2}}, \label{sqcon51}  \\
 	\cos x &=& \pm \frac{1}{\sqrt{2}} \sin z,  \label{sqcon52}\\
 	\cos y &=& \pm \frac{1}{\sqrt{2}}. \label{sqcon53}
\end{eqnarray}
Using Eq.~(\ref{sqcon53}), we obtain $\sin y = \pm1/\sqrt{2}$.  Substituting this condition to Eq.~(\ref{sqcon51}), we obtain $\sin x=\pm 1$ which implies $\cos x = 0$.  Combining this and Eq.~(\ref{sqcon52}), $\sin z= 0$ is satisfied. This leads to the same conditions as Eq.~(\ref{con5}).

Let us check the inverse.  Using Eq.~(\ref{con5}), we can quickly confirm that $g=p=u=0$ and that $f=\pm1/\sqrt{2}$.

\paragraph*{Condition 6.} Because $p - q = 0$ and $u - v =0$, then
\begin{equation*}
 	\cos x \cos y = 0\quad\text{and}\quad\sin y \cos 2z =0,
\end{equation*}
which means that $\cos x = \cos 2z =0$ or $\cos y = \cos 2z = 0$.  
\subparagraph{6-a. $\cos x = \cos 2z =0$.}
In this case, we obtain 
\begin{equation*}
 	u = v = - \sin^2 x \cos^2 y. 
\end{equation*}
Since $u=v=0$, we must have that $\sin x \cos y = 0$.  Since $\sin x \neq 0$, it must be true that $\cos y = 0$.  However, substituting $\cos x = \cos y =0$ to $f$ and $g$ gives $f=g=0$ which contradicts the statement of the condition.
\subparagraph{6-b. $\cos y = \cos 2z =0.$}
In this case, we obtain 
\begin{equation*}
 	u = v =\cos^2 x. 
\end{equation*}
Since $u=v=0$, we have $\sin x\cos x = 0$.  We also have $\sin x \neq 0$, which implies $\cos x = 0$.   On the other hand,
\begin{equation*}
 	p = q = \frac{1}{2} \sin x \sin^2 y \sin 2z.
\end{equation*}
Since $p=q=0$ and $\sin x,\sin y \neq 0$, we obtain $\sin 2z = 0$.  However, we know that $\cos 2z =0$ which contradicts the statement of the condition.   

Then each of aforementioned six conditions are equivalent with each of the following conditions. 
\begin{enumerate}
 	\item $\sin x = 0$.
 	\item $\sin x \neq 0$, $\sin y = 0$.
 	\item $\sin x , \sin y \neq 0$, $\cos x = \cos y = 0$.
 	\item $\cos x = \cos z =0,~  \cos y = \pm 2^{-1/2}$. 
 	\item $\cos x = \sin z =0,~  \cos y = \pm 2^{-1/2}$.
 	\item This condition cannot be satisfied.
\end{enumerate}
Therefore the new form of the necessary and sufficient conditions for implementing the TRH dynamics is as follows.
\begin{enumerate}
 	\item $\sin x = 0$.
 	\item $\sin y = 0$.
 	\item $\cos x = \cos y = 0$.
 	\item $\cos x = \cos z =0,~  \cos y = \pm 2^{-1/2}$.
 	\item $\cos x = \sin z =0,~  \cos y = \pm 2^{-1/2}$.
\end{enumerate}

If a set of parameters $\{x,y,z\}$ satisfies the above five conditions, $\{-x,y,z\}$ also satisfies the same conditions. Since $U(-x,y,z)=U^\dagger(x,y,z)$, this means that if the TRH dynamics ($U^{\rm T}$) of $U$ is implementable, $U^\dagger$ is also implementable.  
In other words, if  
\begin{equation}
	[X_2^{U^\dagger} X_3^{U^{\rm T}}, \tilde{H}_{23}]=[Z_2^{U^\dagger} Z_3^{U^{\rm T}},\tilde{H}_{23}]=0 \label{commAp}
\end{equation}
is satisfied, then 
\begin{equation*}
	[X_2^{U}X_3^{U^\ast},\tilde{H}_{23}]=[Z_2^{U}Z_3^{U^\ast},\tilde{H}_{23}]=0
\end{equation*}
is also satisfied, and vice versa.

The condition given by Eq.~(\ref{commAp}) does hold in general for the AGT scheme. For example, for $x,y,z=\pi/4$, the commutators in Eq.~(\ref{comx}) and Eq.~(\ref{comz}) are non-zero which means that TRH dynamics cannot be implemented by AGT.

To conclude this appendix we give some examples of unitary operations (up to global phase) whose TRH dynamics can be implemented by AGT. For the conditions 1 and 2, 
\begin{equation*}
 	U(x,y,z) = \cos x I - i \sin x Y.
\end{equation*}
For condition 3,
\begin{equation*}
 	U(x,y,z) = \cos z X + \sin z Z.
\end{equation*}
For condition 4, 
\begin{equation*}
	U(x,y,z) = \frac{1}{\sqrt{2}} \left( Y \pm Z \right).
\end{equation*}
For condition 5
\begin{equation*}
 	U(x,y,z) = \frac{1}{\sqrt{2}} \left(Y \pm X\right).
\end{equation*}

\section{Conditions for parallelization}
In this appendix, we show the condition given by Eq.~(\ref{commcondini}) is equivalent to Eq.~(\ref{equicommini}).  The equivalence between  Eq.~(\ref{commcondfin}) and Eq.~(\ref{equicommfin}) can be shown following the same steps.

By expanding Eq.~(\ref{commcondini}), we have
\begin{align}
 	& s(\tau)[ X_1  X_2, \tilde{H}_{12} ] X_3^{W^{(1)}}X_4^{W^{(1)\ast}} X_5^{W^{(2)}} \cdots X_{2L+1}^{W^{(L)}} \nonumber \\
 	& + (1-s(\tau)) X_1 [ X_2 X_3, \tilde{H}_{23}]^{U_3^{(1)}}X_4^{W^{(1)\ast}} X_5^{W^{(2)}} \cdots X_{2L+1}^{W^{(L)}} \nonumber \\
 	&+ s(\tau)   X_1 X_2 [X_3^{W^{(1)}} X_4^{W^{(1)\ast}},\tilde{H}_{34}] X_5^{W^{(2)}} \cdots X_{2L+1}^{W^{(L)}} \nonumber \\
 	&+  (1-s(\tau)) X_1 X_2 X_3^{W^{(1)}} [X_4^{W^{(1)\ast}}  X_5^{W^{(1)}},\tilde{H}_{45}]^{U_5^{(2)}} \cdots X_{2L+1}^{W^{(L)}} \nonumber\\
 	& + \cdots = 0,
\end{align}
where we use the notation $[A_{ij},B_{ij}]^{U_j}:= U_j [A_{ij},B_{ij}]U_j^\dagger$.
Since the commutators in the first two terms are zero, we have
\begin{align*}
 	& s(\tau)   [X_3^{W^{(1)}} X_4^{W^{(1)\ast}},\tilde{H}_{34}] X_5^{W^{(2)}}X_6^{W^{(2)\ast}} \cdots X_{2L+1}^{W^{(L)}} \\
 	&+  (1-s(\tau)) X_3^{W^{(1)}} [X_4^{W^{(1)\ast}}  X_5^{W^{(1)}},\tilde{H}_{45}]^{U_5^{(2)}}X_6^{W^{(2)\ast}} \cdots X_{2L+1}^{W^{(L)}} \\
	&+ s(\tau)   X_3^{W^{(1)}}  X_4^{W^{(1)\ast}} [X_5^{W^{(2)}} X_6^{W^{(2)\ast}},\tilde{H}_{56}] \cdots X_{2L+1}^{W^{(L)}} \\
	& + \cdots = 0.
\end{align*}

By substituting $U^\dagger(x,y,z)= U(-x,y,z)$ in  Eq.~(\ref{comx}), we obtain
\begin{align}
	[X^{U}_2 X_3^{U^{\ast}}, {H_{\rm AGT}^{I}}] =& a(-x,y,z) \cdot (I_2 X_3 - X_2I_3) \nonumber  \\
  	&- b(-x,y,z)\cdot(I_2Z_3 - Z_2 I_3).
 \end{align}
By introducing a trace-less operator defined by 
\begin{equation}
C := b(-x,y,z)Z - a(-x,y,z) X, 
\end{equation}
the commutation relation is transformed to
\begin{align*}
 	& s(\tau) ( C_3^{(3)} I_4 - I_3 C_4^{(3)} )  X_5^{W^{(2)}}X_6^{W^{(2)\ast}}  \cdots X_{2L+1}^{W^{(L)}} \\
 	&+  (1-s(\tau))X_3^{W^{(1)}} ( C_4^{(4)} I_5 - I_4 C_5^{(4)U^{(2)}} )X_6^{W^{(2)\ast}}  \cdots X_{2L+1}^{W^{(L)}} \\
 	&+   s(\tau) X_3^{W^{(1)}} X_4^{W^{(1)\ast}} ( C_5^{(5)} I_6 - I_5 C_6^{(5)} ) \cdots X_{2L+1}^{W^{(L)}} \\
 	&+ \cdots = 0.
\end{align*}
Note that for the third qubit system $\mathcal{H}_3$, only the first term of the above equation has an element containing the $I$ operator.   Thus the first term is linearly independent of the other terms.  Therefore the equation is equivalent to $C^{(3)} =0$ plus the following equation given by
\begin{align*}
 &  (1-s(\tau))( C_4^{(4)} I_5 - I_4 C_5^{(4)U^{(2)}} ) X_6^{W^{(2)\ast}}X_7^{W^{(3)}} \cdots X_{2L+1}^{W^{(L)}} \\
 &+ s(\tau)X_4^{W^{(1)\ast}} ( C_5^{(5)} I_6 - I_5 C_6^{(5)} )X_7^{W^{(3)}} \cdots X_{2L+1}^{W^{(L)}} \\
 &+ (1-s(\tau))X_4^{W^{(1)\ast}} X_5^{W^{(2)}} ( C_6^{(6)} I_7 - I_6 C_7^{(6)U^{(3)}} ) \cdots X_{2L+1}^{W^{(L)}} \\
 &+ \cdots = 0.
\end{align*}
Similarly, we obtain $C^{(4)} = 0$ which reduces the above expression further.  Repeating this reduction, we find that Eq.~(\ref{commcondini}) is equivalent to $C^{(j)}=0$ for all $j$.  Since $C^{(j+1)} = 0$ is equivalent to 
\begin{align*}
	[X_{j+1}^{W^{(\lfloor j/2 \rfloor)}}X_{j+2}^{W^{(\lfloor j/2 \rfloor)\ast}},\tilde{H}_{j+1 ~ j+2}] = 0,
\end{align*}
the equivalence between Eq.~(\ref{commcondini}) and Eq.~(\ref{equicommini}) is shown.

\end{document}